\documentclass{aa}
\usepackage[varg]{txfonts}
\usepackage{amsmath}
\usepackage{amssymb}
\usepackage{graphicx}
\usepackage{inputenc}
\usepackage{subcaption}
\usepackage{chngcntr}

\usepackage[usenames,dvipsnames]{color}

\def\f#1   {Fig.~\ref{#1}}
\def\s#1   {Sec.~\ref{#1}}
\def\tab#1   {Tab.~\ref{#1}}
\def\eq#1   {Eq.~\ref{#1}}
\def\t#1   {Tab.~\ref{#1}}
\newcommand{\smo}{Smol\v{c}i\'{c} }
\newcommand{\zivez}{\v{Z}. Ivezi\'{c} }

\newcommand{\e}{\mathrm{e}}
\usepackage{wrapfig}

\newcommand{\der}{\mathrm{d}}

\newcommand{\lb}{\left(}
\newcommand{\rb}{\right)}
\newcommand{\lbb}{\left[}
\newcommand{\rbb}{\right]}
\newcommand{\lbbb}{\left\{ }
\newcommand{\rbbb}{\right\} }

\newcommand*\segfuncttwo[5]{
#1 = 
\begin{cases}
{#2} &\quad, \quad {#3}\\
{#4} &\quad, \quad {#5}
\end{cases}   }

\newcommand*\segfuncttwoalter[5]{
#1 = 
\begin{cases}
{#2} &\ , \quad {#3}\\
{#4} &\ , \quad {#5}
\end{cases}   }

\title{The XXL survey LII : The evolution of radio AGN luminosity function determined via
parametric methods from GMRT, ATCA, VLA and Cambridge interferometer observations }
\titlerunning{The XXL Survey. XX}
\author{B. \v{S}laus\inst{1}
        \thanks{Corresponding author: \emph{bslaus.phy@pmf.hr}
                },
         V. \smo\inst{1}
         \thanks{Corresponding author: \emph{vs.phy@pmf.hr}
                }
        , \zivez \inst{2}, S. Fotopoulou\inst{3,4}, C. J. Willott\inst{5}, P. Pendo, C. Vignali\inst{6,7}, L. Chiappetti\inst{8}, M. Pierre\inst{9}
}
\authorrunning{B. \v{S}laus}
\institute{Department of Physics, Faculty of Science, University of Zagreb,
Bijeni\v{c}ka cesta 32, 10000 Zagreb, Croatia   \and
University of Washington, Department of Astronomy, PAB 357, 3910 15th Ave NE, Seattle, Washington\and
Center for Extragalactic Astronomy, Department of Physics, Durham University, South Road, Durham DH1 3LE, UK\and
HH Wills Physics Laboratory, School of Physics, University of Bristol, Tyndall Avenue, Bristol, BS8 1TL, UK \and
National Research Council of Canada, Herzberg Astronomy $\&$ Astrophysics Research Centre, 5071 West Saanich Road, Victoria, BC, V9E 2E7, Canada \and Dipartimento di Fisica e Astronomia "Augusto Righi", Università degli Studi di Bologna, Via P. Gobetti 93/2, 40129 Bologna, Italy \and INAF-OAS, Osservatorio di Astrofisica e Scienza dello Spazio di Bologna, Via Gobetti 93/3, 40129 Bologna, Italy \and INAF, IASF Milano, via Corti 12, 20133 Milano, Italy \and Université Paris-Saclay, Université Paris Cité, CEA, CNRS, AIM, F-91191, Gif-sur-Yvette, France} 

\begin{document}

\date{Received ?? Accepted ??}          
\abstract{We model the evolution of active galactic nuclei by constructing their radio luminosity functions. We use a set of surveys of varying area and depth, namely the deep COSMOS survey of $1,916$ AGN sources, the wide shallow 3CRR, 7C and 6CE surveys, containing together $356$ AGNs, and the intermediate XXL-North and South fields consisting of $899$ and $1,484$ sources, respectively. We also used the CENSORS, BRL, Wall $\&$ Peacock and Config surveys, consisting respectively of $150$, $178$, $233$ and $230$ sources. Together, these surveys numbered $5,446$ AGN sources and constrained the luminosity functions at high redshift and over a wide range of luminosities (up to $z \approx 3$ and $\log (L / \mathrm{W Hz^{-1}}) \in [22,29])$. We concentrate on parametric methods within the Bayesian framework and show that the luminosity-dependent density evolution (LDDE) model fits the data best, with evidence ratios varying from "strong" ($>10$) to "decisive" ($>100$) according to the Jeffreys interpretation. \iffalse The median values of the model parameters, defined via $\Phi(L,z) = \Phi_0 \times \lbb  {(1 + z_c)^{p_1}+(1 + z_c)^{p_2} }\rbb / \lbb{\lb \frac{1+z_c}{1+z}   \rb^{p_1}  + \lb \frac{1+z_c}{1+z}   \rb^{p_2} } \rbb$ with $z_c = z_c^*$ for $L > L_a$ and $z_c^* \cdot \lb \frac{ L}{L_a}\rb^a$ for $L < L_a$ equaled $z^* = 2.01$, $L_a = 27.94$, $a = 0.40$, $p_1 = -0.39$, $p_2 = 4.53$, $\Phi^* = -3.92$, $L^* = 22.26$, $\alpha = 1.39$ and $\sigma = 1.40$.\fi  We determine the number density, luminosity density and kinetic luminosity density as a function of redshift, and observe a flattening of these functions at higher redshifts, not present in simpler models, which we explain by our use of the LDDE model. Finally, we divide our sample into subsets according to the stellar mass of the host galaxies in order to investigate a possible bimodality in evolution. We found a difference in LF shape and evolution between these subsets. All together, these findings point to a physical picture where the evolution and density of AGN cannot be explained well by simple models but require more complex models either via AGN sub-populations where the total AGN sample is divided into subsamples according to various properties such as, for example, optical properties and stellar mass, or via luminosity-dependent functions. }             

\maketitle      
\makeatother

\section{Introduction\label{sec:intro}}
\label{sec:Int}

The evolution of active galactic nuclei (AGN) describes their change, in either number or properties, through cosmic time. It is widely accepted that the evolution of AGNs relates closely to the evolution of their host galaxies via AGN feedback (e.g., \citealt{Heckman_Best2014}). The presence of feedback can be deduced both directly via galactic winds (e.g., \citealt{Nesvadba2008}, \citealt{Feruglio2010}, \citealt{Veilleux2013}, \citealt{Tombesi2015}) and X-ray cavities in galaxy clusters (\citealt{Clarke1997}, \citealt{Rafferty2006}, \citealt{McNamaraNulsen2007}, \citealt{Fabian2012}, \citealt{Nawaz2014}, \citealt{Kolokythas2015}), or indirectly as correlations between galactic properties and the mass of its central supermassive black hole (\citealt{Magorrian1998}, \citealt{Ferrarese_Merritt2000}, \citealt{Gebhardt2000}, \citealt{Graham2011},  \citealt{Sani2011}, \citealt{Beifiori2012},  \citealt{McConnell_Ma2013}). It is also a component of the semi-analytic models (e.g. \citealt{Croton2016}, \citealt{Harrison2018}). If we concentrate on AGNs observable in radio wavelengths, statistical analysis of radio-AGN feedback is also possible via luminosity functions (LFs) by estimating the kinetic luminosity or the energy stored in the lobes (e.g. \citealt{Smolcic2017c}, \citealt{Ceraj2018}).

In this work we examine AGNs, observable in the radio part of the spectrum. Of special interest is the tendency throughout the literature to examine specific sub-populations of radio-AGNs, as a possible difference in evolution between these sub-populations could provide further insight into the details of the processes taking place within them. The exact classification, however, varies across the literature, where the division is performed either via relative excess of radio emission, compared to the emission in the optical part of the spectrum, into radio loud (RL) and radio quiet (RQ) AGN (e.g., \citealt{Padovani2015}), via emission lines in the optical spectrum into high or low excitation radio galaxies (HERGs and LERGs, respectively; e.g., \citealt{Pracy2016}, \citealt{Butler2019}, hereafter XXL Paper XXXVI), or a number of other definitions (see \citealt{Padovani2017} for a review of AGN classification). A physical model that could explain the need for AGN sub-populations assumes the existence of two modes of AGN black hole accretion (e.g. \citealt{Heckman_Best2014} for a review) resulting in two distinct populations of AGN: radiatively efficent and radiatively inefficient populations. The radiatively efficient population accretes cold matter onto the central black hole at high Eddington ratios, $\lambda_{Edd}$, of $1\%$ to $10\%$ (\citealt{Heckman_Best2014}, \citealt{Smolcic2017a}, \citealt{Padovani2017}). Here, the Eddington ratio is defined as the bolometric luminosity of the source divided by the maximum possible luminosity due to accretion arising from gravitational force, $\lambda_{Edd} = L_{Bol} / L_{Edd}$, where $L_{Edd} = 1.3 \cdot 10^{38} (M/M_{\odot})\ \mathrm{erg/s}$. According to theory, this population accretes matter via optically thick geometrically thin disk accretion flow (\citealt{Shakura1973}). The radiatively inefficient population accretes hot intergalactic medium at lower Eddington ratios, of typically $\lambda_{Edd} \lesssim 1\% $ (\citealt{Heckman_Best2014}). The physics of accretion is explained theoretically with a geometrically thick optically thin accretion flow (\citealt{Narayan1998}).

A clear way of probing the possible differences in evolution of AGN sub-populations is to construct the radio LFs of the AGN sample, either by relying on the non-parametric methods (e.g. \citealt{Waddington2001}, \citealt{Sadler2007}, \citealt{Donoso2009}, \citealt{Rigby2015}) or modeling the LFs with a functional form decribing their shape and evolution (e.g. \citealt{Smolcic2009}, \citealt{Willott2001}, \citealt{Pracy2016}). The observed trend resulting from such surveys is that there exists a difference in the AGN evolution as a function of AGN luminosity. The space density of the high-luminosity AGN population (luminosities larger than $\log (L / \mathrm{W Hz^{-1}}) \approx 24$) exhibits a strong evolution with redshift up to $z\approx 2$ after which a cut-off is observed (\citealt{Dunlop1990}, \citealt{Willott2001}, \citealt{Pracy2016}). On the other hand the low-luminosity AGNs exhibit little evolution (\citealt{Clewley2004}, \citealt{Smolcic2009}) and the cut-off if it exists occurs at larger redshifts.

In this work we model the LFs of AGNs, using a composite set of surveys of varying area and depth. Since deep surveys constrain the LFs at high redshifts, and large shallow surveys constrain the high luminosity end of the LFs, by combining these types of fields with intermediate fields of medium area and depth, it is possible to robustly model the LFs across a wide range of redshifts and luminosities. A composite set of such a large number of fields, reaching such a depth in redshift, has not yet been analysed. The methodology of parameter estimation and model selection is performed within the Bayesian framework.  

The paper is organized as follows. In Sect \ref{sec:Data} we describe the individual surveys comprising the data used in this work. Sect. \ref{sect:AGNSample} describes threshold imposed to obtain a pure AGN sample. In Sect. \ref{sect:BayesIn} we describe the methodology of Bayesian model fitting and model selection, while Sect. \ref{sect:VmaxCh} describes the complementary method of maximum volumes. Sect. \ref{sect:Models} lists all the examined LF models. In Sect. \ref{sec:REs} and \ref{sect:Discuss} we show the results and discuss them in the context of other publications. Sect. \ref{sec:Sum} gives the summary and conclusion of this work. Throughout this paper we use a cosmology defined with $H_0 = 70 \ \mathrm{km s^{-1} Mpc ^{-1}} $, $\Omega_m = 0.3$, and $\Omega_{\Lambda} = 0.7$. The spectral index, $\alpha$, was defined using the convention in which the radio emission is described as a power law, $S_{\nu} \propto \nu^{\alpha}$, where $\nu$ denotes the frequency, while $S_{\nu}$ is the flux density.

\section{Data}
\label{sec:Data}

In order to maximize the coverage of the L-z plane and sample size, we used a set of radio surveys with varying sizes and depth. Since sources with high radio luminosity are rare, in order to assure a large enough quantity, it is necessary to utilise surveys with large observation areas, such as the $7$C, $6$CE and $3$CRR surveys (\citealt{Willott2001}). Faint sources, on the other hand, are optimally observed via deep surveys, such as the COSMOS survey (\citealt{Smolcic2017a}). We also use intermediate area and depth radio surveys, namely the XXL North and South fields (\citet{SmolcicSlaus2018} hereafter XXL Paper XXIX, \citealt{Butler2017}, hereafter XXL Paper XVIII), in order to bridge the gap between deep and shallow surveys. Additionally we used the CENSORS, BRL, Wall $\&$ Peacock and Config surveys, containing a large percentage of spectroscopic redshifts.

All of the radio catalogues used in this work ($7$C, $6$CE, $3$CRR, XXL-North, XXL-South, COSMOS, listed in Table \ref{tab:fields}) are observed at radio wavelengths. However, the exact frequency varied across surveys. In order to make the datasets more coherent we recalculated all the fluxes to a common frequency of $1400 \ \mathrm{MHz}$ assuming a standard power law shape of radio emission flux $S_{\nu} \propto \nu^{\alpha}$, where $\nu$ denotes the frequency, $S_{\nu}$ the flux density, and $\alpha$ is the spectral index. The value of the spectral index is taken from the corresponding catalogue when it exists, or set to the mean value of that catalogue, as provided by the corresponding publications. The effect of the spectral index on the results is discussed later in Sect. \ref{Sect:Checks}.

\begin{table*}[]
\caption{The surveys used in the estimation of the luminosity functions.  }
\centering
\begin{tabular}{|p{3.2cm}||p{1.5cm}|p{1.5cm}|p{2.5cm}|p{1.5cm}|p{1.8cm}|}
     \hline
      Survey & Area$[\mathrm{deg}^2]$ & Original frequency $[\mathrm{MHz}]$ & Detection limit at  $1400 \ \mathrm{MHz}$ $[\mathrm{m Jy \ beam^{-1}}]$ & Number of sources (AGN)  &  Mean Alpha (AGN)\\
     \hline
     7C & $72.22$ & $151$ & $105$ & $128$  & $-0.64 \pm 0.27$ \\
     6CE  & $338.13$  & $151$ & $421$ & $58$ & $-0.51 \pm 0.32 $\\
     3CRR & $13886.3$ & $178$ & $2,609$ & $170$   & $-0.67 \pm 0.24$ \\
     XXL-North (Inner) & $6.3$  & $610$ & $1.0 $ & $292$ &  $-0.42 \pm 0.49 $\\
     XXL-North (Outer) & $14.2$  & $610$ & $1.0 $  & $607$  &  $-0.48 \pm 0.57 $\\
     XXL-South & $25$& $2100$ & $1.0$  & $1484$  &  $ -0.63 \pm 0.37 $\\
     COSMOS & $2$   & $3000$ & $1.15 \cdot 10^{-2}$ & $1916$   & $-0.80 \pm 0.44$ \\
     CENSORS & $6$   & $1400$ & $7.2$ & $150$   & $-0.80 $ \\
     BRL & $13, 123$   & $408$ & $2,109$ & $178$   & $-0.81 \pm 0.25$ \\
     Wall\ $\&$ \ Peacock & $32, 204$   & $2700$ & $3,167$ & $233$   & $-0.57 \pm 0.55$ \\
     Config & $4, 925$   & $1400$ & $1,300$ & $230$   & $-0.56 \pm 0.37$ \\
 \hline
\end{tabular}
\label{tab:fields}
\end{table*}

\subsection{7C, 6CE and 3CRR}
A set of three wide shallow surveys were the $7$C, $6$CE and $3$CRR fields, observed with the Cambridge Low-Frequency Synthesis Telescope and the Cambridge Interferometer. See \citet{Willott2001} and references therein for details about the surveys, which we briefly summarize below.

The $7$C field was observed at frequencies of $151\  \mathrm{MHz}$, detecting sources above $0.5\  \mathrm{Jy}$ (\citealt{Willott2001}). The complete area of the survey, which consists of three distinct regions: $7$C-I, $7$C-II and $7$C-III, equalls $72.22 \ \mathrm{deg}^2$ (i.e. $0.022 \ \mathrm{sr}$). The spectral indices were determined using multifrequency radio data for $7$C-I, $7$C-II, and $38 \ \mathrm{MHz}$ $8C$ data for $7$C-III (\citealt{Lacy1999}), resulting altogether in a mean spectral index of $\alpha \approx -0.64$. The redshift information was derived from follow-up optical and near-infrared observations (\citealt{Willott2001}). Most of the redshift were determined spectroscopically ($\approx 85\%$) while the remaining sources have photometric redshifts. The number of sources in the $7$C catalogue equals $128$.

The $6$CE survey at $151\ \mathrm{MHz}$ covered an area of $\approx 340\ \mathrm{deg}^2$ ($0.103 \ \mathrm{sr}$) capturing sources with flux density $2 \ \mathrm{Jy} < S_{151\ \mathrm{MHz}} < 3.93 \ \mathrm{Jy}$ (\citealt{RawlingsEales2001}). The catalogue contained $59$ sources, of which all but three have spectroscopically determined redshifts. The mean spectral index, obtained by a polynomial fit to the multi-frequency data (\citealt{RawlingsEales2001}) equaled $\alpha \approx -0.51$. For more details on the catalogue see \citet{RawlingsEales2001}.

The $3$CRR catalogue, from observations at $178\ \mathrm{MHz}$ spans an area of $\approx 13, 900\ \mathrm{deg}^2$ ($4.23  \ \mathrm{sr}$), with a detection limit of $10.9 \ \mathrm{Jy}$. The redshift information is present for all $173$ sources in the sample as describd in \citet{Willott1999}. The spectral index, calaculated at rest frame $151\ \mathrm{MHz}$ had a mean of $\alpha \approx -0.67$.

The $1.4 \ \mathrm{GHz}$ rest-frame radio luminosities for the $7$C, $6$CE and $3$CRR used here were computed from flux, redshift and spectral index values given in the corresponding catalogs, using a newer cosmology defined in Sec. \ref{sec:Int} ($H_0 = 70 \ \mathrm{km s^{-1} Mpc ^{-1}} $, $\Omega_m = 0.3$, and $\Omega_{\Lambda} = 0.7$).

\subsection{XXL-North}
\label{sec:radio}

The observations of the XXL-North field were performed at $610\  \mathrm{MHz}$ with the Giant Metrewave Radio Telescope (GMRT). The observations were divided into two distinct parts: the inner part of the field spanned an area of $11.9 \ \mathrm{deg}^2$ (the XMM-Large Scale Structure, XMM-LSS field). The data were taken from an earlier study by \citet{Tasse2007}, and then re-reduced as discussed in \citet{Slaus2020} (hereafter XXL Paper XLI). These observations reach a mean rms of $200\  \mathrm{\mu Jy \ beam^{-1}}$. The remaining $18.5 \ \mathrm{deg}^2$, were observed by XXL Paper XXIX. They have a mean rms of $45\  \mathrm{\mu Jy \ beam^{-1}}$. The number of observed sources in both parts of the field was $5,434$, using a signal-to-noise ratio of $S/N\geq 7$. The spectral indices for both parts of the field were obtained by matching the catalogue with another radio catalogue, the NRAO Very Large Array Sky Survey (NVSS) at $1400\  \mathrm{MHz}$ (\citealt{Condon1998}) as described in XXL Paper XXIX. The mean spectral index equaled $-0.65$ for the inner part of the field, and $-0.75$ for the outer, due to the difference in survey depth. The $157$ multi-component sources were created from components as described in XXL Paper XLI. More details on the observations and the corresponding catalog can be found in XXL Paper XXIX.

In order to obtain the source redshifts, the catalogue was cross-matched with a multiwavelength catalogue from \citet{Fotopoulou2016}, hereafter XXL Paper VI, using only the subset of the catalog that has identifications in the Spitzer Infrared Array Camera (IRAC) Channel 1 band at $3.6 \ \mathrm{\mu m}$ (PI M. Bremer, limiting magnitude of 21.5 AB), to obtain uniform depth. The redshifts of the IRAC-detected sources were determined photometrically using the full multi-wavelength data (Fotopoulou, in prep.). Details on the redshift accuracy can be found in XXL Paper XLI. The IRAC survey covered roughly $80 \%$ of the radio field. After further removing the noisy edges of the radio map, the area of the inner part of the field equaled $6.3 \ \mathrm{deg}^2$ and the area of the outer $14.2 \ \mathrm{deg}^2$.

\subsection{XXL-South}

The $25  \ \mathrm{deg}^2$ of the XXL-South field were observed with the Australia Telescope Compact Array (ATCA), at $2.1 \ \mathrm{GHz}$ (XXL Paper XVIII). The observations reached a depth of $\approx 41 \  \mathrm{\mu Jy \ beam^{-1}}$. The details of the observations are described in XXL Paper XVIII. The catalogue consists of $6,239$ single component sources and an aditional $48$ sources composed of multiple components. The spectral indices were determined by matching the $2.1 \ \mathrm{GHz}$ catalogue with the Sydney University Molonglo Sky Survey (SUMSS) at $843 \ \mathrm{MHz}$ (\citealt{Bock1999}) reaching sources with peak flux density of $6 \ \mathrm{m Jy}$. After taking into account the bias that arises from a high detection limit of the SUMSS survey, the median spectral index was estimated at $\alpha \approx -0.75$.

The catalogue was furthermore matched with a multiwavelength catalogue using a likelihood technique as described in \citet{Ciliegi2018} (hereafter XXL Paper XXVI). The multiwavelength catalogue contained data from near-infrared and optical up to X-ray data (for details see XXL Paper VI). The cross-correlation process resulted in $4,770$ optical/NIR counterparts, with $414$ of them also detected in the X-ray band (XXL Paper XXVI). Since $12$ of these sources were classified as stars, they were removed from the sample, resulting in a catalogue of $4,758$ sources (\citealt{Butler2018}, hereafter XXL Paper XXXI). There are $1,110$ spectroscopic redshifts and $3,648$ photometric redshifts listed in the catalogue (XXL Paper XXXI, XXL Paper VI). The details concerning the accuracy of the photometric redshifts and the overall redshift distribution of the sample can be found in XXL Paper XXXI. The median spectral index of the matched catalogue is flatter and equals $-0.45$ (XXL Paper XXXI).

\subsection{COSMOS}

The deepest radio survey used in this works is the VLA-COSMOS $3 \ \mathrm{GHz}$ Large Project (\citealt{Smolcic2017a}). The area of the observed field also covered by multiwavelength data, is $2 \ \mathrm{deg}^2$ and the detection limit at $5 \sigma$ equaled $11.5 \  \mathrm{\mu Jy \ beam^{-1}}$. The full catalogue contained $10,830$ sources of which $67$ are multi-component. The spectral indices were derived from cross-correlation with the $1.4 \ \mathrm{GHz}$ Joint catalogue by \citet{Schinnerer2010}, using survival analysis to account for the bias of different detection limits, as described in \citet{Smolcic2017a}. The mean spectral index was estimated at $\alpha = -0.73$. 

The radio catalogue was further matched with a multiwavelength catalogue as described in \citet{Smolcic2017b}. This resulted in $\approx 93 \%$ of the sources obtaining a counterpart ($8035/8696$ in the unmasked part of the field). For $7778$ of these sources there exists a redshift estimate. Of these, $2740$ are spectroscopic ($\approx 34 \%$), while the other $5123$ are photometric. For details on the redshift estimation see \citet{Delvecchio2017} and \citet{Smolcic2017b}.

\subsection{CENSORS}

Another study used in this work is the Combined EIS–NVSS Survey Of Radio Sources (CENSORS) by \citet{Brookes2008}, containing $150$ radio sources. The catalogue comes from the spectroscopic observations of the $1.4 \ \mathrm{GHz}$ sources observed by \citet{Best2003}. The area of observations equalled $6 \ \mathrm{deg}^2$, and the detection limit reached $7.2 \ \mathrm{mJy}$. The catalogue contained spectroscopic redshifts for $71 \%$ of the sample, while the remaining redshifts were determined via the K-z relation, as described by \citet{Brookes2008}. The spectral indices were lacking and were set to a constant value of $-0.8$, following the assumptions made in \citet{Brookes2008}. Four sources from the sample whose radio emission comes from star formations were classified by \citet{Brookes2008}. The rest of the sample consists of AGN sources.

\subsection{The BRL Sample}

The Best, Röttgering $\&$ Lehnert (BRL) sample is defined from the Molonglo Reference Catalogue, with the spectroscopic data compiled and observed by \citet{Best1999}, \citet{Best2000}, \citet{Best2003} and \citet{Best2003b}. The radio observations were performed at $408  \ \mathrm{MHz}$, with a detection limit of $5 \ \mathrm{Jy}$. The area of observations was bounded by declination $\delta \in [-30^o, 10^o]$, excluding the Galactic plane $|b| > 10^o$ 

\subsection{The Wall $\&$ Peacock Sample}

The sample by \citet{Wall1985} covers an area of $\approx 32, 000 \ \mathrm{deg}^2$ ($9.81 \ \mathrm{sr}$). The observations were performed via ANRAO/Parkes, NRAO/Greenbank, and MPIfR/Bonn. The frequency of observations equalled $2.7 \ \mathrm{GHz}$, with a detection limit of $2 \ \mathrm{Jy}$ at this frequency. The catalogue consists of $233$ sources, with $171$ of them ($73 \ \%$) having measured redshifts. The remaining redshifts were estimated from $V$-band magnitudes, as described by \citet{Wall1985}. The spectral indices of the sources were determined from additional $5 \ \mathrm{GHz}$ data. Given the high detection limit of the survey, we assumed it consists purely of AGNs.

\subsection{CoNFIG}

The Combined NVSS-FIRST Galaxies (CoNFIG) sample (\citealt{Gendre2008}) comes from observations at $1.4 \ \mathrm{GHz}$, selecting NRAO Very Large Array (VLA) Sky Survey (NVSS) sources from the north field of the Faint Images of the Radio Sky at Twenty centimetres (FIRST) survey, spanning $\approx 5000  \ \mathrm{deg}^2$ ($1.5 \ \mathrm{sr}$). The catalogue contains $274$ sources, selected above a detection limit of $1.3 \ \mathrm{Jy}$. Redshifts are determined for $89 \%$ of the sample. The redshifts are mosltly spectroscopic ($230$ sources), and some determined via the $R-z$ relation ($14$ sources), as described by \citet{Gendre2008}. The spectral indices present in the catalogue were determined from observations at lower frequencies, up to $178 \ \mathrm{MHz}$.

\section{AGN Samples}
\label{sect:AGNSample}
In order to obtain a pure AGN sample, we selected a further sub-sample of the above described catalogues. For the shallow fields this was not required as the detection limit of these surveys was very bright. This ensured that the observed sources were AGN. 

For the XXL-North survey, following the source counts from \citet{Smolcic2017a}, XXL Paper XLI set an additional threshold leaving only sources with flux $> 1 \ \mathrm{mJy}$. This threshold ensures the sample consists purely of AGN at all fluxes. Here we decided to prioritize the purity of our sample. Although this removes a number of sources from the analysis, namely the fainter part of the sample, this is not a problem since that part of the sample is constrained well with the COSMOS catalogue used also in this work to constrain the LFs. The number of sources in the inner part of the field equaled $292$ and in the outer $607$. The mean spectral indices were $-0.42$ and $-0.48$ for the inner and outer parts of the field, respectively.

A similar procedure was performed for the XXL-South survey. Although detailed classification of sources into AGNs and star forming galaxies (SFGs) can be found in XXL Paper XXXI, since in constraining the luminosity functions we also use a deeper COSMOS survey covering fainter sources, we could impose again a conservative threshold leaving only sources with $> 1 \ \mathrm{mJy}$. In analogy with the XXL-North field, this threshold ensures the resulting sample consists purely of AGNs. The number of AGNs in our sample thus equaled $1,484$. Of these sources $\approx 24\%$ have spectroscopic redshifts, and the mean spectral index of the AGN sample equaled $-0.63$.

For the COSMOS field we used only sources with excess radio emission relative to that expected from the galaxy's star formation rate as described in \citet{Smolcic2017b}. The AGN sample was defined via a ratio of radio emission compared to the star formation rate obtained from the infrared emission (computed via SED fitting) as described by \citet{Delvecchio2017}. The number of sources in the final AGN sample used for this work thus equaled $1,916$, and the mean spectral index was $-0.80$. Of these sources, $\approx 32 \%$ have spectroscopic redshifts.

\begin{figure*}
\centering
\includegraphics[width=0.7\textwidth]{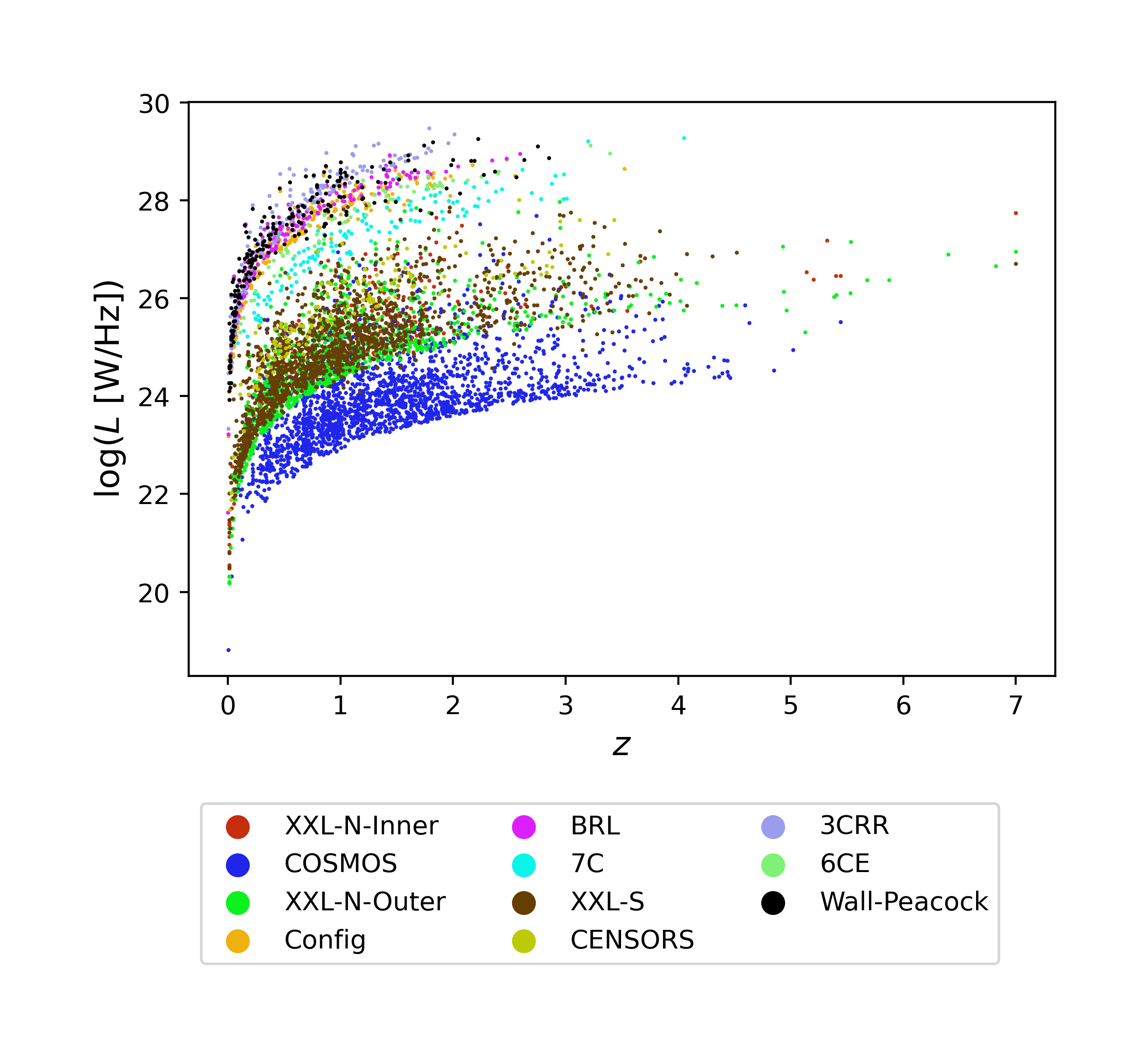}
\caption{ The redshift-luminosity plot of the complete composite sample, of radio AGNs used in this work. The names of the fields are denoted in the legend.}
\label{fig:RedLum}
\end{figure*}

The complete AGN sample,consisting of $5,446$ sources, is shown in Fig. \ref{fig:RedLum} as a redshift-luminosity plot, which illustrates visually the ranges in redshift and luminosity that each survey spans.
As visible from the plot, redshifts reach $z \approx 4$, while luminosities span approximately $\log (L / \mathrm{W Hz^{-1}}) \in [22,29]$. Outside of these limits, the results should be interpreted carefully.

\section{Bayesian modeling of luminosity functions}
\label{sect:BayesIn}
The luminosity functions in this work were modeled within the Bayesian framework. The aim of Bayesian modeling is to determine the posterior $P(\Theta | D, M)$, or the probability density function of the model parameters $\Theta$, given $D,M$ which represent the data and model respectively. The posterior is calculated using the prior $\pi$ and the likelihood $\mathcal{L}$ (\citealt{Thrane2019}):
\begin{equation}
P(\Theta | D, M) = \frac{\pi(\Theta) \mathcal{L}(D|\Theta)}{E}
\end{equation}
where $E$ is the normalisation factor also called the evidence:
\begin{equation}
E = \int \pi(\Theta) \mathcal{L}(D|\Theta) \der \Theta
\end{equation}
The likelihood function describes the measurements and we discuss it in detail in the next subsection. The prior function quantifies our knowledge of the parameters, before any measurements are taken (\citealt{Thrane2019}). In this work the priors were chosen to be uniform, reflecting no prior assumptions about the model parameters. Priors for parameters expressed as logarithms were taken to be uniform in the logarithmic scale. In order to perform the numerical calculations, we used the "\textsc{Dynesty}" program package by \cite{Speagle2020}, which uses dynamic nested sampling (\citealt{Skilling2004}, \citealt{Higson2019}).

\subsection{Likelihood function}
A crucial step in the process of Bayesian parameter estimation is determining the likelihood function. We followed here \citet{Marshall1983} (see also \citealt{Christlein2009} for a more detailed derivation). By dividing the complete luminosity-redshift space into infinitesimal cells $\der z \der L$ and assuming that each cell is small enough to contain up to one source, invoking the Poisson distribution, the probability of observing $N$ sources of the complete sample is:
\begin{equation}
p = \prod_i^N \lambda_i \e^{-\lambda_i} \cdot \prod_j \e^{-\lambda_j}
\end{equation}
where $\lambda$ is the expected number of sources per bin. The first product goes over the complete sample of $N$ sources, while the second one takes into account that all the remaining cells must remain empty. The expected density of sources in a given luminosity bin $\der L$ is given by the luminosity function $\phi(z,L)$:
\begin{equation}
\lambda = \phi\ \der V \der L = \phi \frac{\der V}{\der z} \der z \der L
\end{equation} 
By taking the customary logarithm of the probability and rearranging the sums we obtain: 
\begin{equation}
\ln (p) = \sum_i \ln \lb \phi_i \frac{\der V_i}{\der z} \der z \der L \rb - \int  \phi \frac{\der V}{\der z} \der z \der L
\label{logLikeNoErr}
\end{equation}
where $\phi_i$ and $V_i$ are associated with a particular source of the catalogue. The first sum goes over the observed sample. The second sum, which was turned into an integral, covers the whole available $(z,L)$ space. The limits of the integral therefore follow the detection limit of the survey, i.e. it numbers all the cells where in principle a source could be observed. It therefore also follows that the integral equals the total predicted number of sources above the detection limit of the survey (\citealt{Christlein2009}). The log-Likelihood, $\ln \mathcal{L}$, is defined as (e.g. \citealt{Marshall1983}):
\begin{equation}
\ln \mathcal{L} = -2 \ln (p) = - 2\sum_i \ln \lb \phi_i \frac{\der V_i}{\der z} \der z \der L \rb + 2 \int  \phi \frac{\der V}{\der z} \der z \der L
\end{equation}
The expression can be further simplified by noting that not all terms depend on the luminosity function parameters. The first term of the last equation can be divided into:
\begin{equation}
\sum_i \ln \lb \phi_i \frac{\der V_i}{\der z} \der z \der L \rb = \sum_i \ln \phi_i +  \sum_i \ln \lb \frac{\der V_i}{\der z} \der z \der L \rb
\label{omit1}
\end{equation}
The second term of this relation does not depend on the luminosity function parameters and, as such, provides the $\ln L$ relation with a constant value not important in the minimisation process. It can therefore be omitted. We have finally:
\begin{equation}
\ln \mathcal{L}  = - 2\sum_i \ln \phi_i + 2 \int  \phi \frac{\der V}{\der z} \der z \der L
\label{omit2}
\end{equation}
This expression is the one found commonly in the literature (e.g. \citealt{Kelly2008}, \citealt{Yuan2020}). Furthermore, this expression can be generalised naturally to multiple fields $j$ with different detection limits and observational areas as:
\begin{equation}
\ln \mathcal{L}  = - 2   \sum_{i,j} \ln \phi_i  + 2 \sum_j  \int_j \phi \frac{\der V}{\der z} \der z \der L
\label{omit2}
\end{equation}
where the first sum covers all the sources from all the composite fields and each integral in the second sum reaches the depth of the corresponding field as denoted by the lower limit. If the incompleteness of the survey near the detection limit is significant, it can be included as a separate completeness function, as described in the next subsection.

\subsection{Completeness corrections}
\label{LFCorr}

The completeness corrections of each survey can be introduced naturally by using a smooth detection limit which is a function of flux, instead of an abrupt cutoff. The corrections for each survey were taken from their respective papers, as described below.

For the XXL-North field the correction is given in XXL Paper XXIX. This correction corresponds to the one arising from noise near the detection limit. In XXL Paper XLI we also introduced another correction arising from the losses during the matching of radio data with the multiwavelength catalogue which were not negligible. This correction is a function of redshift.

For the COSMOS field, the correction can be found in \citet{Smolcic2017a} in their Figure 16 or Table 02. Finally, as seen from XXL Paper XVIII and \citet{Willott2001} the other catalogues can be considered complete. Therefore for these catalogues no corrections were included.

\subsection{Model comparison}
\label{Sect:AIC_BIC}
A feature of the Bayesian formalism discussed above is the ability to compare the fit between different models. A direct comparison is obtained by calculating the odds ratio (\citealt{Thrane2019}):
\begin{equation}
O_{21} = \frac{p(M_2|D,I)}{p(M_1|D,I)} = \frac{E(M_2)p(M_2|I)}{E(M_1)p(M_1|I)}
\end{equation}
where we have re-introduced the evidence from Sect. \ref{sect:BayesIn}, $E = p (D|M,I)$ (\citealt{Liddle2007}). If, furthermore, there is no model preferred by the priors, the odds ratio reduces to a ratio of evidences called the Bayes factor:
\begin{equation}
B_{21} = \frac{E(M_2)}{E(M_1)}
\end{equation}
This ratio is also commonly expressed as a difference in log-scale. It should be noted that this method naturally selects models with a lower number of parameters, or in other words it incorporates Occam's razor (\citealt{Liddle2007}).

Another, more approximative, method is the Akaike information criterion (AIC; \citealt{Akaike1974}, \citealt{Liddle2007}, XXL Paper VI). The definition of the AIC value is:
\begin{equation}
AIC = 2 k - 2 \ln \mathcal{L}
\end{equation}
where $k$ denotes the number of parameters. The model with the lower value of AIC is the one that corresponds to a better fit. It can be seen that this method also penalises a larger number of parameters. In other words, the Occam's razor is included naturally within the comparison. 

Lastly we have the Bayesian information criterion (BIC; \citealt{Schwarz1978}, \citealt{Liddle2007}, XXL Paper VI), similar to AIC, and defined as:
\begin{equation}
BIC = k \ln N - 2 \ln \mathcal{L}
\end{equation}
where $N$ is the number of data points. This value is the numerical approximation for the Bayes factor. From the above expression it is also immediately clear that for a sufficiently large $N$, the BIC penalty for a large number of parameters is stronger than that of AIC.

\section{Maximum volume method}
\label{sect:VmaxCh}

A complementary non-parametric method to the Bayesian formalism is the method of maximum volumes described by \citet{Schmidt1968} (see also \citealt{Felten1976}, \citealt{Avni1980}, \citealt{Page2000} and \citealt{Yuan2013}, \citealt{Novak2017}). This method incorporates naturally the inherent bias arising from detection limits of observations by taking into account that more luminous sources are detectable from farther away. The luminosity function values in each luminosity and redshift bin are estimated by summing the inverse maximum volumes of possible observation for each source: $1/V_{Max,i}$. The errorbars of the data points are estimated assuming Gaussian statistics (\citealt{Marshall1985}, \citealt{Boyle1988}, \citealt{Page2000}, \citealt{Novak2017}), except when the number of sources equaled less than $10$. In those situations we used the values calculated by \citet{Gehrels1986}. Furthermore, the number of sources used here was an effective number, determined from maximum volumes, as described in \citet{Ananna2022}.

A further complication arises from the fact that we are using multiple fields with varying depth. In order to coherently determine the value of the luminosity function, we follow the procedure described in \citet{Avni1980} (see also \citealt{Giallongo2005}, \citealt{Russell2011}, \citealt{Gruppioni2013} XXL Paper VI). The maximum volume for each source was calculated by taking into consideration all the fields where in principle this source could have been detected. For a given range of redshifts $[z_1, z_2]$ we write:
\begin{equation}
V_{Max,\ i} = \sum_j \omega_j \int_{z_1}^{z_{Up}(i,j)} \frac{\der V}{\der z} \der z
\end{equation}
Here the sum goes over all the fields $j$ where source $i$ could have been observed, and $\omega_j$ denotes the area of the respective field. The upper limit of the integral was the minimum between $z_2$ and the maximum redshift of possible detection given the detection limit of the corresponding survey. In other words, we take into account that sources from the shallow fields are detectable in all the deeper fields as well, which modifies the value of their maximum volume (\citealt{Gruppioni2013}). In practice the integral in the last equation was calculated numerically by dividing the redshift interval into smaller subsets, following the procedure described in XXL Paper XLI. The detection limits of each survey were shifted to $1400 \ \mathrm{MHz}$ by assuming a power law and using the mean value of the spectral index.

\section{Luminosity function models}
\label{sect:Models}
One of the main aims of this work is to use the Bayesian framework to compare between different luminosity function models. We list here all the different models used in this work. The complete list of models is also summarized in Table \ref{tab:models}.

\subsection{Local luminosity function}
The shape of the local luminosity function is usually described by a power-law with an exponential cut-off (\citealt{Saunders1990}, \citealt{Sadler2002}, \citealt{Smolcic2009}):
\begin{equation}
\label{SadLocal}
\Phi_0(L) = \Phi^* \lb \frac{L}{L^*} \rb^{1-\alpha} \exp \lbbb  \frac{-1}{2\sigma^2} \lbb \log \lb 1 + \frac{L}{L^*} \rb \rbb^2 \rbbb
\end{equation}
where the base of the logarithm is $10$. Here $L^*$ is the break luminosity, $\Phi^* $ the normalisation and $\sigma$ the high-luminosity slope. Another possible choice would have been the double power law used often for radio and X-ray AGN samples (\citealt{Dunlop1990}, \citealt{Mauch2007}, \citealt{Smolcic2017c}, XXL Paper VI). We discuss our choice in more detail in the discussion. Lastly, another possibility is the bimodal model from \citet{Willott2001} which has a different form for the low and high luminosity end of the sample. This model is discussed separately in Sect. \ref{Sect:BimodalModel} below.

\subsection{Evolution}
The evolution of the aforementioned local LFs is given by both density and luminosity evolution as (e.g. \citealt{Smolcic2017c}): 
\begin{equation}
\label{Evol:PDEPLE}
\Phi(L,z) = (1 + z) ^{\alpha_D} \times \Phi_0 \lbb \frac{L}{(1+z)^{\alpha_L}} \rbb
\end{equation}
where $\alpha_D$ and $\alpha_L$ are the parameters quantifying density and luminosity evolution respectively. In this work we refer to this model as Sadler$+02$. We also tested pure density evolution (PDE) and pure luminosity evolution (PLE) where only one evolution parameter is different from zero. Another parametrization of luminosity function evolution can be found in \citet{Novak2017}:
\begin{equation}
\Phi(L,z) = (1 + z) ^{(\alpha_D + z \beta_D)} \times \Phi_0 \lbb \frac{L}{(1+z)^{(\alpha_L + z \beta_L)}} \rbb
\label{eq:novakmodel}
\end{equation}
where $\beta$ parameters quantify the change of evolution with redshift. We refer to this model as Novak$+18$. In order to account for the difference in evolution between the high and low luminosity end of the sample, we also investigated the luminosity dependent density evolution (LDDE; \citealt{Schmidt1983}, \citealt{Ueda2003}) following XXL Paper VI:
\begin{equation}
\Phi(L,z) = \Phi_0 \times \frac{(1 + z_c)^{p_1}+(1 + z_c)^{p_2} }{\lb \frac{1+z_c}{1+z}   \rb^{p_1}  + \lb \frac{1+z_c}{1+z}   \rb^{p_2} }
\label{eq:LDDE}
\end{equation}
where:
\begin{equation}
\segfuncttwo{z_c }{z_c^*}{L > L_a}{z_c^* \cdot \lb \frac{ L}{L_a}\rb^a}{L \leq L_a}
\label{eq:La}
\end{equation}
where $L_a$ is the luminosity where the evolution changes according to relation (\ref{eq:La}), $z_c$ redshift after which the evolution changes and $p_{1,2}$ the parameters of evolution. Lastly, as introduced by \citet{Massardi2010} and \citet{Bonato2017}, the evolution of sources can be described by a luminosity dependent luminosity evolution model (LDLE). Where the break luminosity $L^*$ evolves as:
\begin{equation}
L^*(z) = L^*(0) \cdot 10 ** \left\{  k_{evo} z \left[  2z_{top} - 2z^{m_{ev}} z_{top}^{1-m_{ev}} / \left(1 + m_{ev}\right)   \right] \right\}
\label{eq:LDLE_Lstar}
\end{equation}
and:
\begin{equation}
z_{top} = z_{top,0} + \frac{\delta z_{top}}{1 + L^*(0)/L}
\label{eq:LDLE_ztop}
\end{equation}
Here $k_{evo}$ and $m_{ev}$ are free parameters of evolution, while $z_{top,0}$ and $\delta z_{top}$ determine the redshift where the evolution changes.

\subsection{Bimodal luminosity function model}
\label{Sect:BimodalModel}
A special case for both the local luminosity function and its evolution is the bimodal model taken from \citet{Willott2001}. We refer to it in this work as Willott$+01$. The shape and evolution of the sample have a different analytical form for the high and low luminosity end of the sample. Following \citet{Smolcic2009}, in this work we used model "C" from \citet{Willott2001}, being the most flexible one, defined as:
\begin{equation}
\Phi = \Phi_l + \Phi_h
\end{equation}
where $\Phi_l$ is the low luminosity end of the function:
\begin{equation}
\segfuncttwoalter{ \Phi_l }{  \Phi_{l0} \lb \frac{L}{L_l^*} \rb^{-\alpha_l} \exp \lb \frac{-L}{L_l^*}  \rb (1 +z )^{k_l}  }{ z < z_{l0} }{\Phi_{l0} \lb \frac{L}{L_l^*} \rb^{-\alpha_l} \exp \lb \frac{-L}{L_l^*}  \rb (1 +z_{l0} )^{k_l} }{z > z_{l0} }
\end{equation}
and $\Phi_h$ is the high luminosity end:
\begin{equation}
\segfuncttwo{ \Phi_h }{  \Phi_{h0} \lb \frac{L}{L_h^*} \rb^{-\alpha_h} \exp \lb \frac{L_h^*}{-L}  \rb \cdot \exp \lbb \frac{-1}{2}  \lb \frac{z - z_{h0}}{z_{h1}}  \rb \rbb  }{z < z_{h0} }{\Phi_{h0} \lb \frac{L}{L_h^*} \rb^{-\alpha_h} \exp \lb \frac{L_h^*}{-L}  \rb  \cdot \exp \lbb \frac{-1}{2}  \lb \frac{z - z_{h0}}{z_{h2}}  \rb \rbb  }{z > z_{h0} }
\end{equation}
Here $L^*$ denotes the break luminosity, $\Phi_0$ the normalisation and $\alpha$ the slope of the luminosity functions. Parameter $z_{0}$ is the redshift at which the evolution changes. These parameters exist separately for the high and low luminosity end of the sample as denoted by the extra indices $h,l$. Parameters $k_l$, $z_{h1}$ and $z_{h2}$ quantify the evolution.

\begin{table}[]
\caption{LF models used in this work, corresponding list of free parameters, and their number $N_{Par}$. }
\centering
\begin{tabular}{|p{1.5cm}||p{5.5cm}|p{0.6cm}|}
     \hline
      Model & Parameters & $N_{Par}$ \\
     \hline
Sadler+02 & $\Phi^*, L^*,\sigma ,\alpha, \alpha_D, \alpha_L$ & 6 \\
 PDE &$\Phi^*, L^*,\sigma ,\alpha, \alpha_D$ & 5 \\
 PLE &$\Phi^*, L^*,\sigma ,\alpha, \alpha_L$ & 5 \\
 Novak+18&$\Phi^*, L^*,\sigma ,\alpha, \alpha_D, \alpha_L,  \beta_D, \beta_L$ & 8 \\
 LDDE & $\Phi^*, L^*,\sigma ,\alpha, z_c^*, a, L_a, p_1, p_2$ & 9 \\
 LDLE & $\Phi^*, L^*,\sigma ,\alpha, k_{evo}, m_{ev}, z_{top,0},\delta z_{top}$ & 8 \\ 
Willott+01 & $\Phi_{l0}, L_l^*, \alpha_l, k_l, z_{l0}, \Phi_{h0}, L_h^*, \alpha_h, z_{h0}, z_{h1}, z_{h2}$ & 11 \\

 \hline
\end{tabular}

\label{tab:models}
\end{table}

\section{Results}
\label{sec:REs}

\subsection{Testing the methodology}

Before using the observed data discussed in Sect. \ref{sec:Data}, we tested the methodology by using simulated data. For this purpose we created custom Python codes which created catalogues of mock-observations starting from an assumed LF. We then tried to re-create the assumed LF by modeling the LFs using our methodology on simulated mock-observations. We tested both the parametric method described in Sect. \ref{sect:BayesIn} and the method of maximum volumes described in Sect. \ref{sect:VmaxCh}. The test were performed for a wide range of different fields and their combinations, starting with a wide range of assumed LFs. The results were always in good agreement with the starting luminosity function, within the range defined by the uncertainties defined via $90\%$ quantiles for a sample of LFs drawn randomly from the posterior. This provided us with confidence that the methodology used in this work is sound. 

As an example, we describe here the process performed on a Schechter LF model, with a superposition of PDE and PLE evolution (named Sadler$+02$ within this work). The area of the field was set to $40.46 \ \mathrm{deg}^2$, and the detection limit to $50 \ \mathrm{\mu Jy}$. This resulted in a simulated catalogue of $6378$ mock sources above the detection limit, created by randomly selecting sources via the assumed LF. The starting parameters of the LF are given in Tab. \ref{Tab:ParametersMock}. The scatter in redshifts was set here to be negligible, but a finite uncertainty in redshifts, via hierarchical bayesian interference, was also tested. The parameter modeling was performed on this simulated data set, using the same codes later used on observational data. The retrieved parameters are shown also in Tab. \ref{Tab:ParametersMock}. The detection limit in this example was a step-function i.e. the completeness corrections were not present. We also assumed a mean spectral index of $-0.7$. The codes were also tested for non-negligible completeness corrections, by introducing completeness correction as a separate function during the integration of log-Likelihood. The methodology was tested for all the models described in Sect. \ref{sect:Models} and on different areas and depths of mock-catalogues. The parameters of the LFs were always retrieved successfully.

\begin{table}[]
\caption{Assumed and retrieved parameters resulting from the modeling of LFs on simulated data. As described in the text, a mock catalogue was created using assumed LF models. This catalogue was then used to model the LFs in order to test the validity of the modeling methodology.}
\centering
\begin{tabular}{|p{1.5cm}||p{1.5cm}|p{1.5cm}|p{1.1cm}|p{1.1cm}|}
     \hline
      Parameter & Assumed & Retrieved & $+2\sigma $ & $-2\sigma $ \\
     \hline
$\log \Phi^*$ & -5.10 & -4.99 & 0.77 & 0.50 \\
$\log L^*$ & 23.0 & 22.82 & 0.83 & 2.40 \\
$\alpha$ & 1.50 & 1.50 & 0.08 & 0.40 \\
$\sigma$ & 1.50 & 1.53 & 0.22 & 0.16 \\
$\alpha_D$ & 1.00 & 0.92 & 0.64 & 0.56 \\
$\alpha_L$ & 0.50 & 0.68 & 0.80 & 0.86 \\
 \hline
\end{tabular}
\label{Tab:ParametersMock}
\end{table}

\subsection{The luminosity functions using the COSMOS, XXL, 3CRR, 7C and 6CE surveys}

The same methods, which were successfully tested on simulations, were used on real observed data, namely the COSMOS, XXL, 3CRR, 7C and 6CE fields, as well as the CENSORS, BRL, Wall $\&$ Peacock and Config surveys, described in Sect. \ref{sec:Data}, combining all the catalogues as a single composite survey. We estimated the best model parameters for all luminosity functions discussed in Sect. \ref{sect:Models}. The numerical calculations were performed by the \textsc{Dynesty} program package (\citealt{Speagle2020}), which resulted in both the model parameter posteriors and the marginal likelihoods for each model. We also obtained the posterior samples useful for plotting the luminosity functions as they preserved the correlation between the parameters of the model. 

According to our data, the best fitting model is the LDDE model of evolution, with an exponential local form, described by relations \ref{SadLocal}, \ref{eq:LDDE} and \ref{eq:La}. The relative standing of each fit was assessed by comparing their marginal likelihoods. Apart from this, we also used the approximate AIC and BIC methods, introduced in Sect. \ref{Sect:AIC_BIC}. The resulting values of the model comparison are listed in Tab. \ref{Tab:ModelCompare}. We show a comparison between the best fitting LDDE model and all of the other models. We list the values of three different methods: difference in logarithm of evidence, AIC and BIC, separately. The relative standing of the models remains the same across all three methods, with the LDDE model consistently being the preffered one. According to the Jeffrey's interpretation of evidence ratios (e.g. \citealt{Kass1995}), the interpretation varies from "strong" ($>10$)  to "decisive" ($>100$) in favour for the LDDE model. 

\begin{figure*}
\centering
\includegraphics[width=0.95\textwidth]{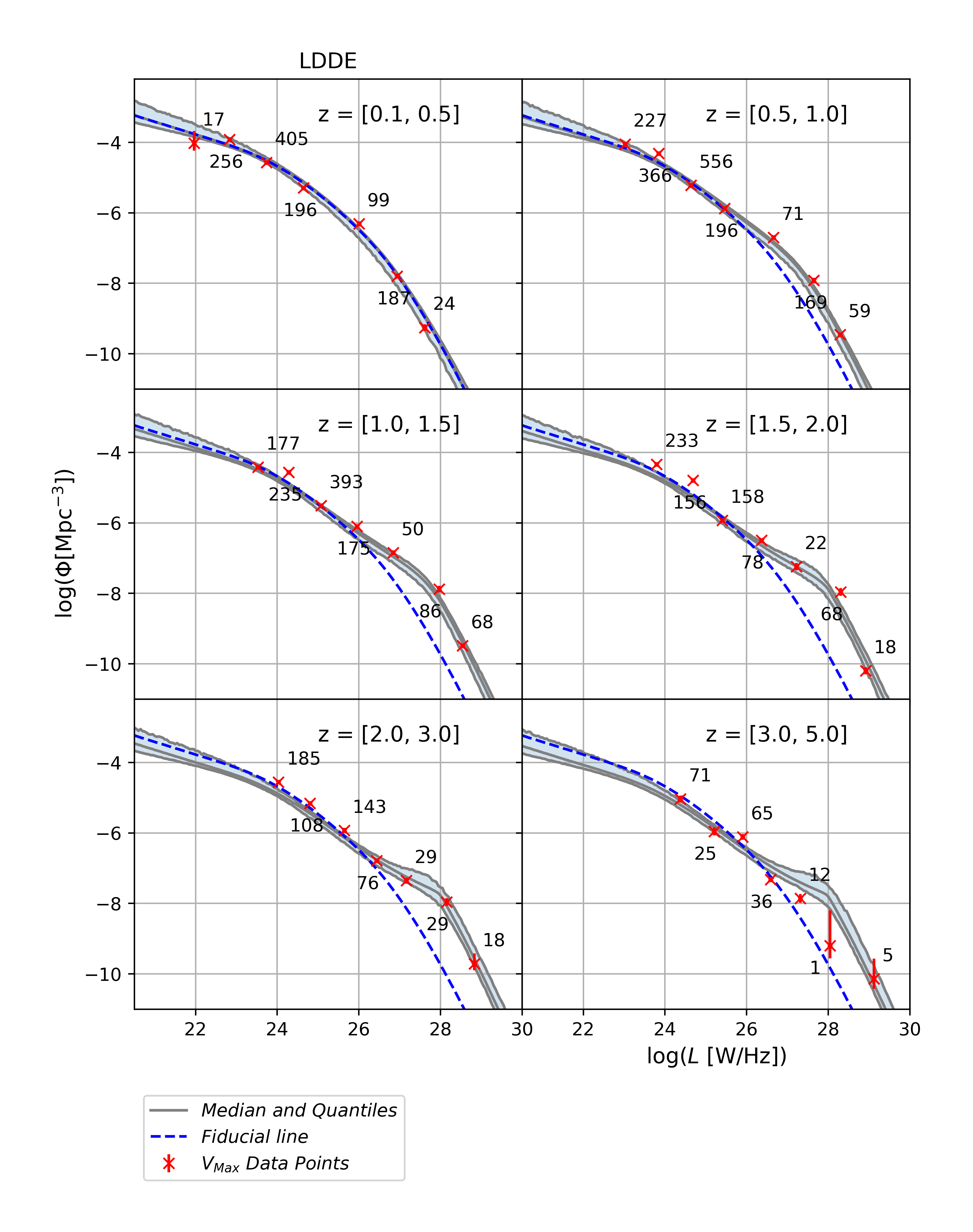}
\centering
\caption{The luminosity functions modeled using using the COSMOS, XXL, 3CRR, 7C and 6CE surveys, obtained by two complementary methods: Bayesian modeling and the method of maximum volumes. Grey lines denote the median and $90\%$ quantiles of the parametric Bayesian inference. These values were obtained by randomly drawing samples from the posterior. The crosses denote the non-parametric method of maximum volumes, together with the corresponding error-bars. The uncertainties were derived assuming Poisson errors, except when the number of sources was lower than $10$. For such data points, the uncertainties were represented by tabulated errors determined by \citet{Gehrels1986}. Here the number of sources was an effective number, determined from maximum volumes, as described in \citet{Ananna2022}. We also show the number of sources creating each data-point. The blue dashed fiducial line denotes the LF determined in the first redshift bin.}
\label{fig:LFs_LDDE}
\end{figure*}

The LDDE model LFs are shown in Fig. \ref{fig:LFs_LDDE} since this was the best fitting model. The grey lines correspond to the median and the $90\%$ quantiles, and were obtained from a random sample of LFs drawn from the posterior. Apart from the Bayesian method, we also show the data points obtained from the maximum volume method described in Sect. \ref{sect:VmaxCh}. It can be seen that the two complementary methods give consistent results. Certain discrepancies between the methods can be seen in the last subplot of the figure, but at such high redshifts ($z>3$) the number of sources is small and the LF less constrained, as visible by the number of sources for each data point given in Fig. \ref{fig:LFs_LDDE}.

\begin{figure*}
\centering
\makebox[\textwidth][c]{\includegraphics[width=1.1\textwidth]{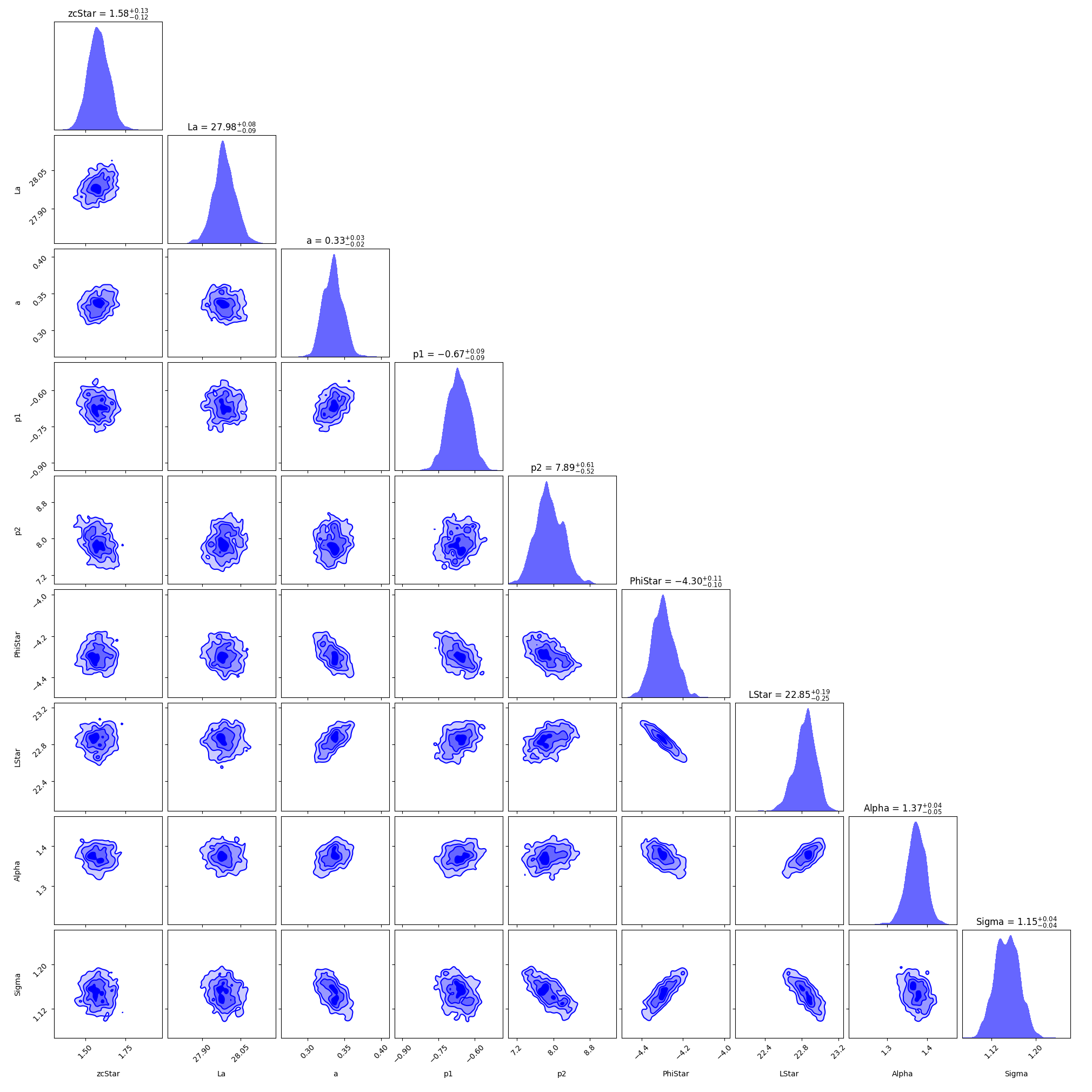}}
\centering
\caption{The Corner-plot showing the posterior distribution of each parameter of the LDDE model. The resulting samples and weights taken from the posterior were further smoothed as described in \citealt{Speagle2020} to obtain the plotted probability density functions. }
\label{fig:CORNER}
\end{figure*}

In Fig. \ref{fig:CORNER} we show the resulting parameters of the LDDE model via corner-plot of the posterior probability density functions. The break luminosity equaled $\log (L^*/ \mathrm{W Hz^{-1}}) = 22.28 ^{ +0.42} _{- 0.55}$ and $z^*= 2.01 ^{+ 0.61} _{-0.38}$. The degeneracy in $p_1$ and $p_2$ parameters is an expected occurrence, as seen from equation \ref{eq:LDDE}, but it was eliminated by choosing the prior so that it encompasses only one peak. The modeling was also tested without this simplification, and the results were qualitatively identical. The resulting values of the parameters determined from the posterior probability density functions are shown within Fig. \ref{fig:CORNER} and listed within Table \ref{Tab:ParameterPosteriors}.

\begin{figure}
\centering
\includegraphics[width=0.5\textwidth]{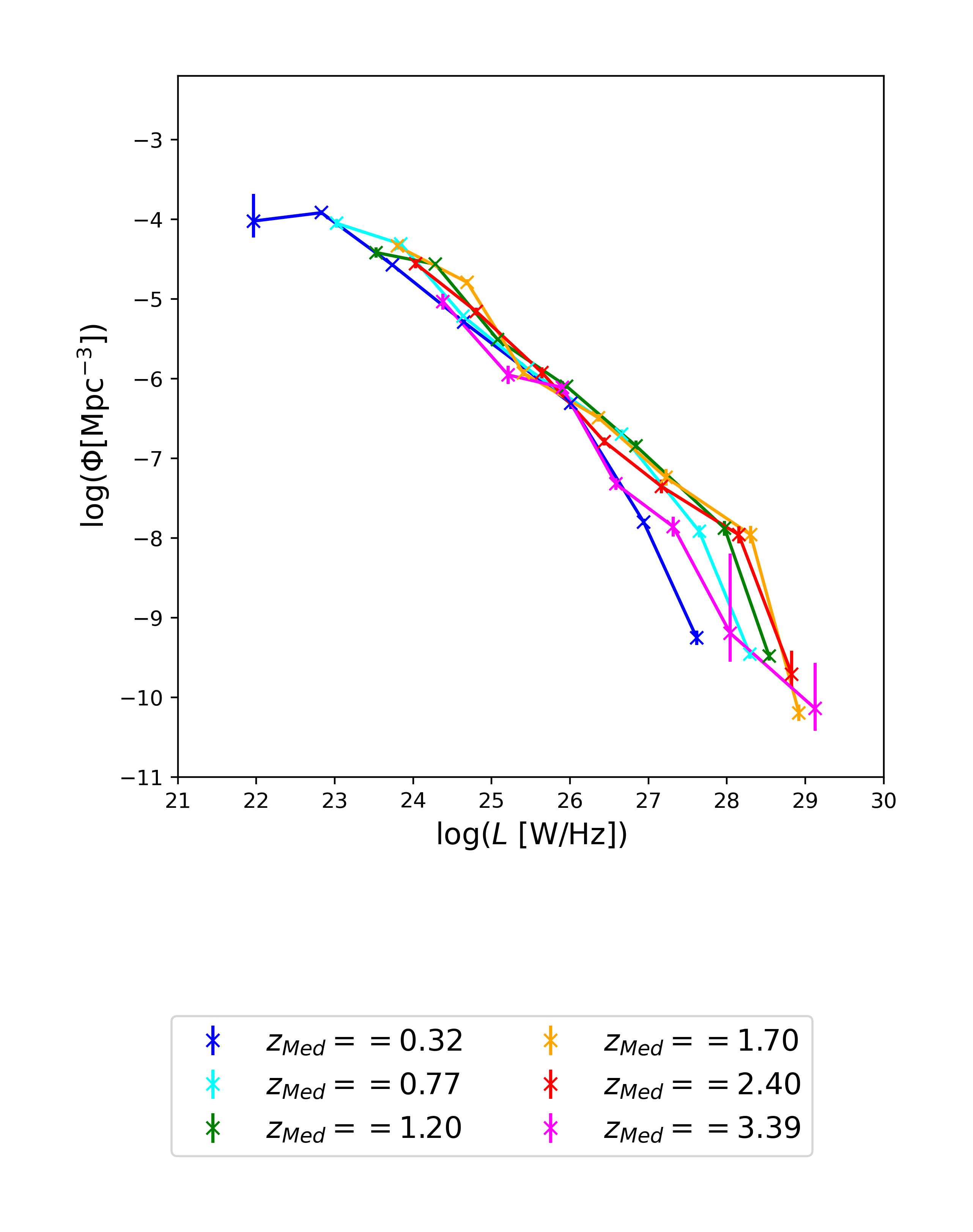}
\centering
\caption{ The non-parametric LFs determined in different redshift bins via method of maximum volumes, as described in the text (Sect. \ref{sect:VmaxCh}), shown overlaid on top of each other, in order to display their evolution. The uncertainties were derived assuming Poisson errors, except when the effective number of sources, determined from maximum volumes, as described in \citet{Ananna2022}, was lower than $10$. For such data points, the uncertainties were represented by tabulated errors determined by \citet{Gehrels1986}. The evolution is stronger for high-luminosity sources. The last redshift bin is created using a smaller subsample of sources and is, as described in the text, less credible.}
\label{fig:VmaxPlot}
\end{figure}

We also show, in Fig. \ref{fig:VmaxPlot}, the LFs created only via the non-parametric method of maximum volumes, as these data points trace the AGN sample dirrectly, without any need of assumed models. The LFs for different redshift bins are shown overlaid together, so the evolution of AGNs is clearly visible. The evolution is stronger for high luminosity sources. The redshift bin with median redshift $z_{Med} = 3.38$ is created using a lower number of sources and the LF is therefore less constrained. The high redshift LF could also point towards a turnover in density at these redshifts, which is less clear from the modeled LFs. However, we note again that the models are not constrained well above $z \approx 3$.

\begin{figure*}
\centering
\includegraphics[width=0.9\textwidth]{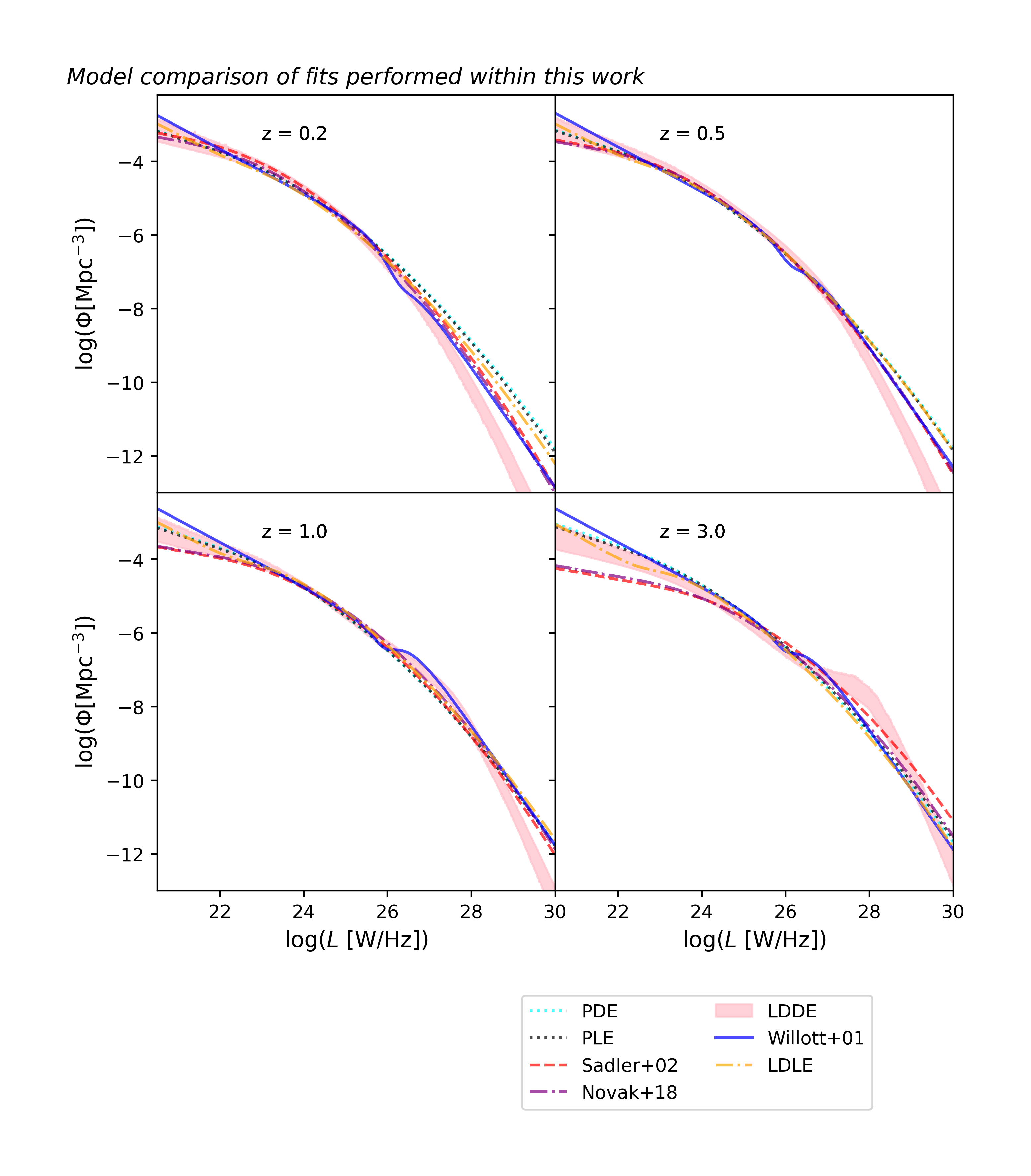}
\centering
\caption{The comparison between the best fitting LDDE model and the other non-optimal fits, all performed within this work on the same composite data set. The shaded area for the LDDE model denotes the $90\%$ quantiles of the parametric Bayesian inference, obtained by randomly drawing samples from the posterior. The other models are represented by medians. }
\label{fig:compareThisWork}
\end{figure*}

\begin{figure*}
\centering
\includegraphics[width=0.9\textwidth]{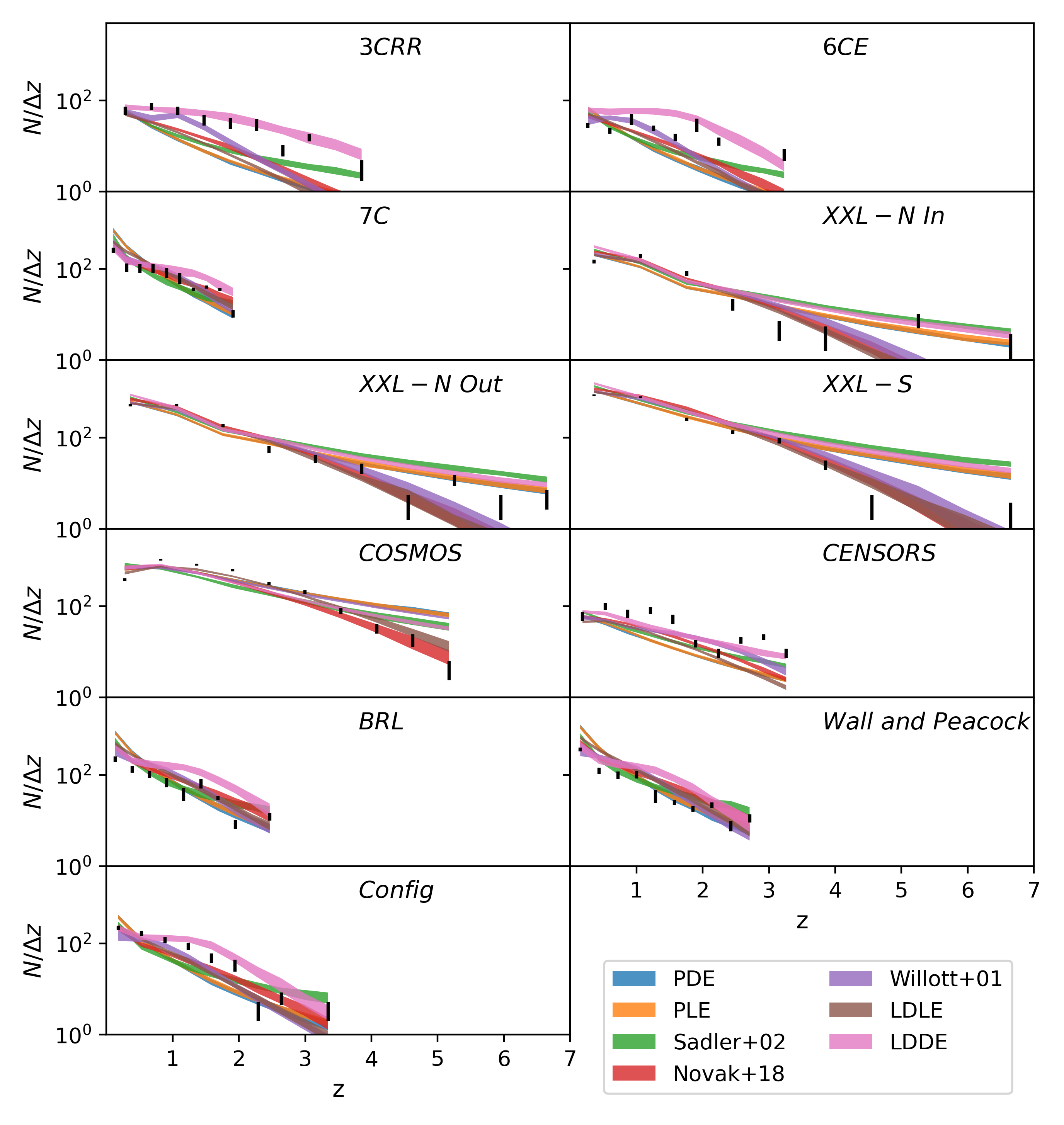}
\centering
\caption{The redshift distribution of sources, shown as a number of sources divided by the width of the redshift bin. The figure shows both the observed redshift histograms as data points, as well as the model prediction lines. The shaded areas are the $85\%$ quantiles, obtained by randomly drawing samples from the posterior. }
\label{fig:RedshiftFit}
\end{figure*}

A comparison between the best fitting LDDE model and the other non-optimal fits performed within this work is shown in Fig. \ref{fig:compareThisWork}. Furthermore, separate plots for each non-optimal fit are given as supplementary material in the appendix. As a further means of model comparison, in Fig. \ref{fig:RedshiftFit} we show the redshift distributions of sources from each survey, compared to the model predictions. The model predictions were obtained by integrating the LF models, taking into account the detection limit of each survey and the corresponding incompleteness, described in Sect. \ref{LFCorr}. The figure shows the model predictions for each LF model used within this work. It can be seen that the LDDE model is the one which is able to reproduce the redshift distributions best, in particular for what concerns the shallow fields of large area, where the other models fail to reproduce the redshift distributions well. This is a consequence of the fact that these models are not flexible enough to describe the "bump" present in the LFs at high luminosities.

\begin{table}[]
\caption{Comparison of the LDDE model with other models using three different methods, as described in the text.  }
\centering
\begin{tabular}{|p{1.5cm}||p{1.8cm}|p{1.8cm}|p{1.8cm}|}
     \hline
      Model & $ 2 \cdot \log B_{21} $ & $-\Delta $AIC & $ -\Delta $BIC \\
     \hline
Sadler+02 & 716.36 & 726.65 & 707.31 \\
 PDE & 1166.54 & 1181.03  & 1155.25 \\
 PLE & 1103.32 & 1117.92 & 1092.14 \\
 Novak+18 & 243.36 & 250.66 & 244.22  \\
Willott+01 & 381.18 & 371.89 & 384.78 \\
LDLE & 477.38 & 483.85 & 477.40 \\
 \hline
\end{tabular}
\label{Tab:ModelCompare}
\end{table}

\begin{table}[]
\caption{Parameters of the best fitting LDDE model. The model is provided in the text in relations (\ref{eq:LDDE} and \ref{eq:La}). The standard deviation, provided by the \textsc{Dynesty} package, is asymmetric.}
\centering
\begin{tabular}{|p{1.5cm}||p{1.8cm}|p{1.8cm}|p{1.8cm}|}
     \hline
      Parameter & Mean & $+2\sigma $ & $-2\sigma $ \\
     \hline
$z_C^*$ & 1.58 & 0.13 & 0.12 \\
$\log L_a$ & 27.98 & 0.08 & 0.09 \\
$a$ & 0.33 & 0.03 & 0.02 \\
$p_1$ & -0.67 & 0.09 & 0.09 \\
$p_2$ & 7.89 & 0.61 & 0.52 \\
$\log \Phi^*$ & -4.30 & 0.11 & 0.10 \\
$\log L^*$ & 22.85 & 0.19 & 0.25 \\
$\alpha$ & 1.37 & 0.04 & 0.05 \\
$\sigma$ & 1.15 & 0.04 & 0.04 \\
 \hline
\end{tabular}
\label{Tab:ParameterPosteriors}
\end{table}

\subsection{Comparison with the literature}

\begin{figure*}
\centering
\begin{subfigure}{.99\textwidth}
\includegraphics[width=0.7\textwidth]{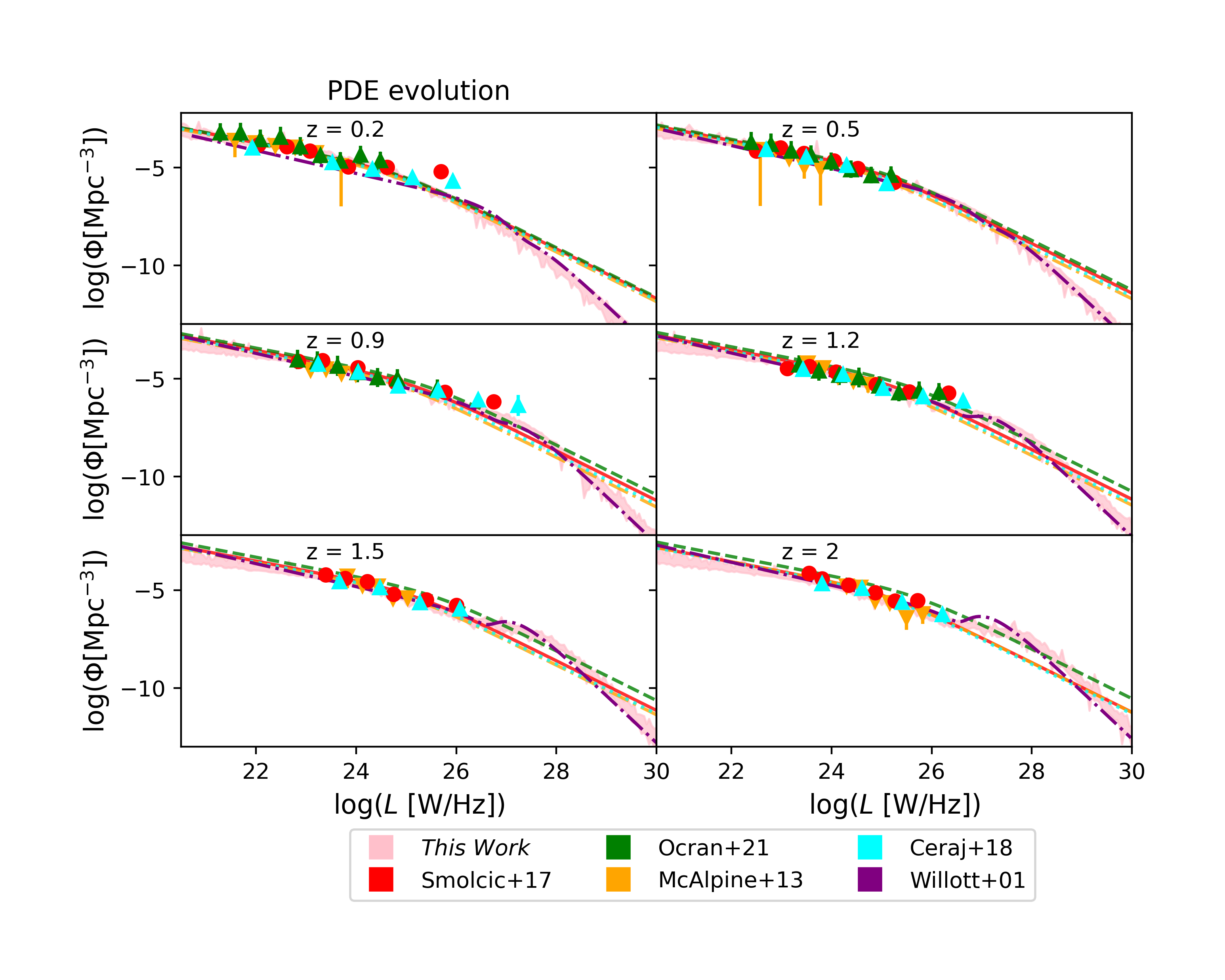}
\centering
\end{subfigure}
\begin{subfigure}{.99\textwidth}
\includegraphics[width=0.7\textwidth]{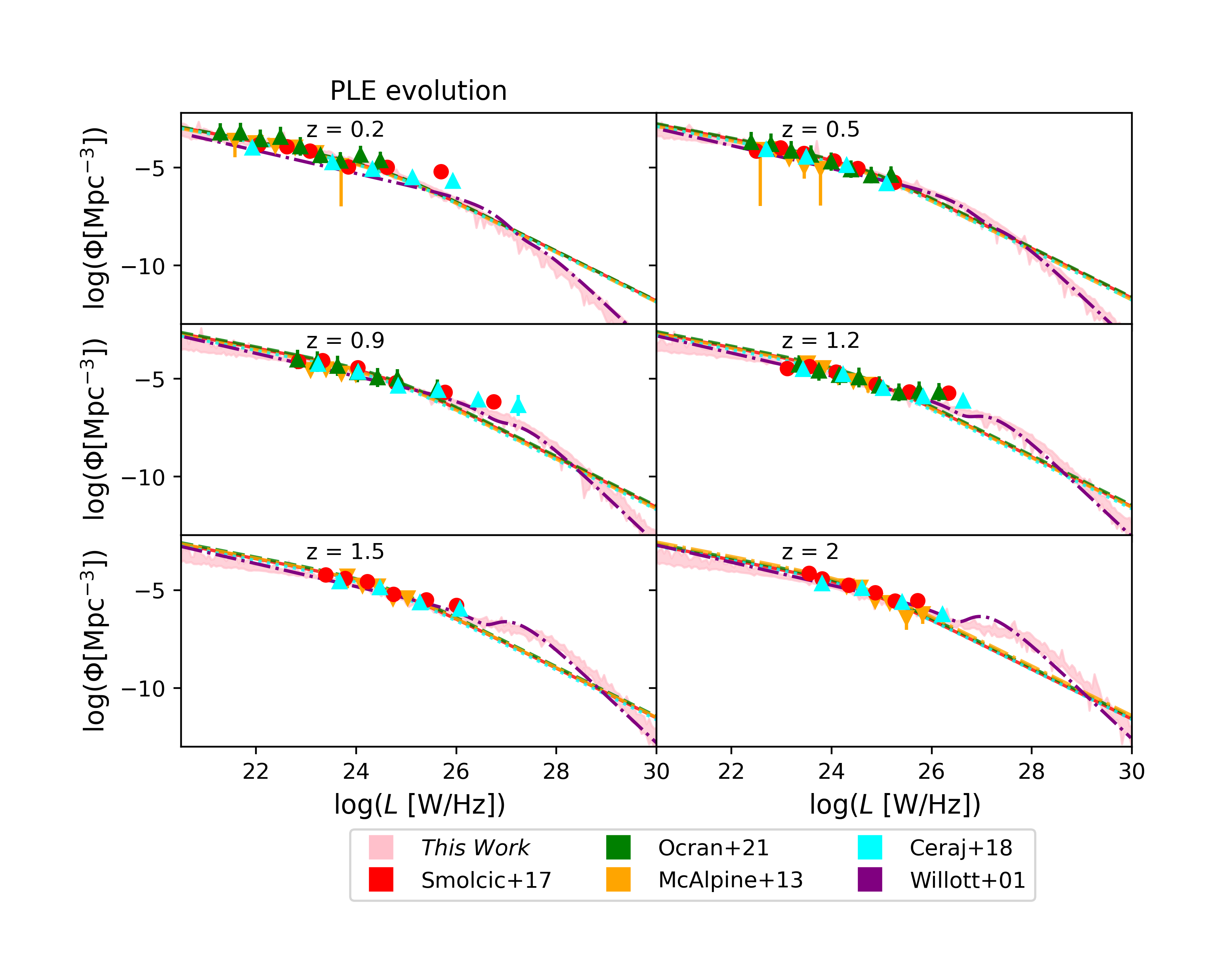}
\centering
\end{subfigure}
\caption{Comparison of our LDDE model with the models from literature, shown separately for PDE and PLE literature evolution models, as denoted above the figures. The used surveys are denoted in the legend. We also show the maximum volume data points taken from the literature, in the same color as the corresponding luminosity function model line. The results of this work, represented by $90\%$ quantiles are given in pink. The Willott LF shown is the one derived by \citet{Willott2001}.}
\label{fig:LFs_comparePDEPLE}
\end{figure*}

The best-fitting LDDE model from this work was compared to a set of radio luminosity function models from the literature, namely \citet{Willott2001}, \citet{McAlpine2013}, \citet{Smolcic2017c}, \citet{Ceraj2018} and \citet{Ocran2021}. The survey from \citet{Willott2001} is already described in Sect. \ref{sec:Data}. The sample from \citet{McAlpine2013} contained $942$ sources observed in the radio with the VLA. The sample from \citet{Smolcic2017c} contained $1800$ AGN sources from the COSMOS field. \citet{Ceraj2018} LFs were determined also from the COSMOS field, with a sample of $1604$ sources. LFs by \citet{Ocran2021} were created from $486$ AGNs from the ELAIS N1 field observed at $610 \ \mathrm{MHz}$. Since the majority of surveys model the luminosity functions with pure PDE or PLE evolution, we show the comparison on two different plots shown in Fig. \ref{fig:LFs_comparePDEPLE}, one for each type of evolution, in order to make them more intelligible. The surveys used in the comparison are listed in the legend. Both plots also show the model from \citealt{Willott2001}, with the parameters taken from the corresponding paper, as it is neither a PDE or a PLE model. As seen from the plots, our results are broadly consistent with earlier surveys, although our comparison based on bayesian evidence comparison shows the LDDE model to be the preferred model (see Tab. \ref{Tab:ModelCompare}). None of the surveys, however, except \citealt{Willott2001}, feature a bump at higher luminosities present in our results. The model from \citealt{Willott2001} shows a difference in evolution as a function of luminosity, but the exact shape of the luminosity functions differs somewhat from our model. There is also a difference in the model from \citet{Willott2001} with other models at lower luminosities but this is a result of their sample not being able to constrain these values well, as discussed in their paper. This is because the sample from \citet{Willott2001} contains $356$ sources mostly from only the higher luminosity set of our sample. Since our whole sample uses a composite set of surveys, our results are not affected by this.

\subsection{Additional checks}
\label{Sect:Checks}
In order to test the robustness of our results, a few additional checks were performed. Firstly we examined the effect of spectral indices on the results. In order to asses this we re-modeled the luminosity functions using different values of spectral indices: a mean value of $-0.7$ for all sources and a mean value of the corresponding field for each source. The quantitative results remain consistent. Secondly, we assessed the redshift uncertainties of the XXL-N field, this being the field with largest uncertainties in redshift. This could have been done by using the hierarchical Bayes method (see \citealt{Loredo2004}, \citealt{Aird2010}, XXL Paper VI). However, since the fields of intermediate depth in this work are already represented by the XXL-S field, a conservative check was performed by simply omitting the XXL-N field. The re-modeling of the LFs again gave consistent results, proving the uncertainties in the redshifts of the XXL-N field do not modify our results. Lastly we note that the same ranking between the evolution models is obtained by using the double power law function by \citet{Dunlop1990} for the local shape of the LF. Overall the results of the model selection seem to be a true consequence of the physical processes within AGN and not a result of unforeseen biases and point towards LDDE model being the best one.

\subsection{Effect of spectral indices on the model parameters}

The luminosities of the sources, and the flux re-calculation from different frequencies, depend on the value of the spectral indices. However, not all of the sources used in the LF modeling had a determined spectral index. As already noted, for the missing indices we used the mean spectral index of the corresponding survey. This does not take into account the  frequency or redshift dependence of the spectral indices (e.g. \citealt{Tisanic2020}). In order to asses the effect of spectral index on the parameters of the best fitting LDDE model, we performed again the model parameter estimation using first only the sources with a determined spectral index, and then assuming a mean spectral index of $-0.7$ for all sources. The results remain consistent, with the LDDE model being determined as the best one. The newly determined model parameters are listed in Tabs. \ref{Tab:ParameterPosteriorsNomeans} and \ref{Tab:ParameterPosteriorsMEAN}.

\begin{table}[]
\caption{Parameters of the best fitting LDDE model for modeling using only the determined values of spectral indices.}
\centering
\begin{tabular}{|p{1.5cm}||p{1.8cm}|p{1.8cm}|p{1.8cm}|}
     \hline
      Parameter & Mean & $+2\sigma $ & $-2\sigma $ \\
     \hline
$z_C^*$ & 1.47 & 0.15 & 0.12 \\
$\log L_a$ & 27.82 & 0.10 & 0.09 \\
$a$ & 0.43 & 0.05 & 0.04 \\
$p_1$ & -0.12 & 0.08 & 0.08 \\
$p_2$ & 7.29 & 0.57 & 0.65 \\
$\log \Phi^*$ & -4.15 & 0.14 & 0.13 \\
$\log L^*$ & 22.31 & 0.39 & 0.58 \\
$\alpha$ & 1.24 & 0.11 & 0.17 \\
$\sigma$ & 1.18 & 0.04 & 0.04 \\
 \hline
\end{tabular}
\label{Tab:ParameterPosteriorsNomeans}
\end{table}

\begin{table}[]
\caption{Parameters of the best fitting LDDE model for modeling using a mean value of spectral index set to $-0.7$.}
\centering
\begin{tabular}{|p{1.5cm}||p{1.8cm}|p{1.8cm}|p{1.8cm}|}
     \hline
      Parameter & Mean & $+2\sigma $ & $-2\sigma $ \\
     \hline
$z_C^*$ & 1.56 & 0.15 & 0.14 \\
$\log L_a$ & 27.96 & 0.11 & 0.09 \\
$a$ & 0.33 & 0.03 & 0.02 \\
$p_1$ & -0.64 & 0.09 & 0.09 \\
$p_2$ & 7.35 & 0.58 & 0.49 \\
$\log \Phi^*$ & -4.19 & 0.18 & 0.12 \\
$\log L^*$ & 22.65 & 0.27 & 0.43 \\
$\alpha$ & 1.36 & 0.06 & 0.09 \\
$\sigma$ & 1.20 & 0.04 & 0.05 \\
 \hline
\end{tabular}
\label{Tab:ParameterPosteriorsMEAN}
\end{table}

\subsection{Number and luminosity density}

\begin{figure}
\centering
\begin{subfigure}{.5\textwidth}
\includegraphics[width=0.8\textwidth]{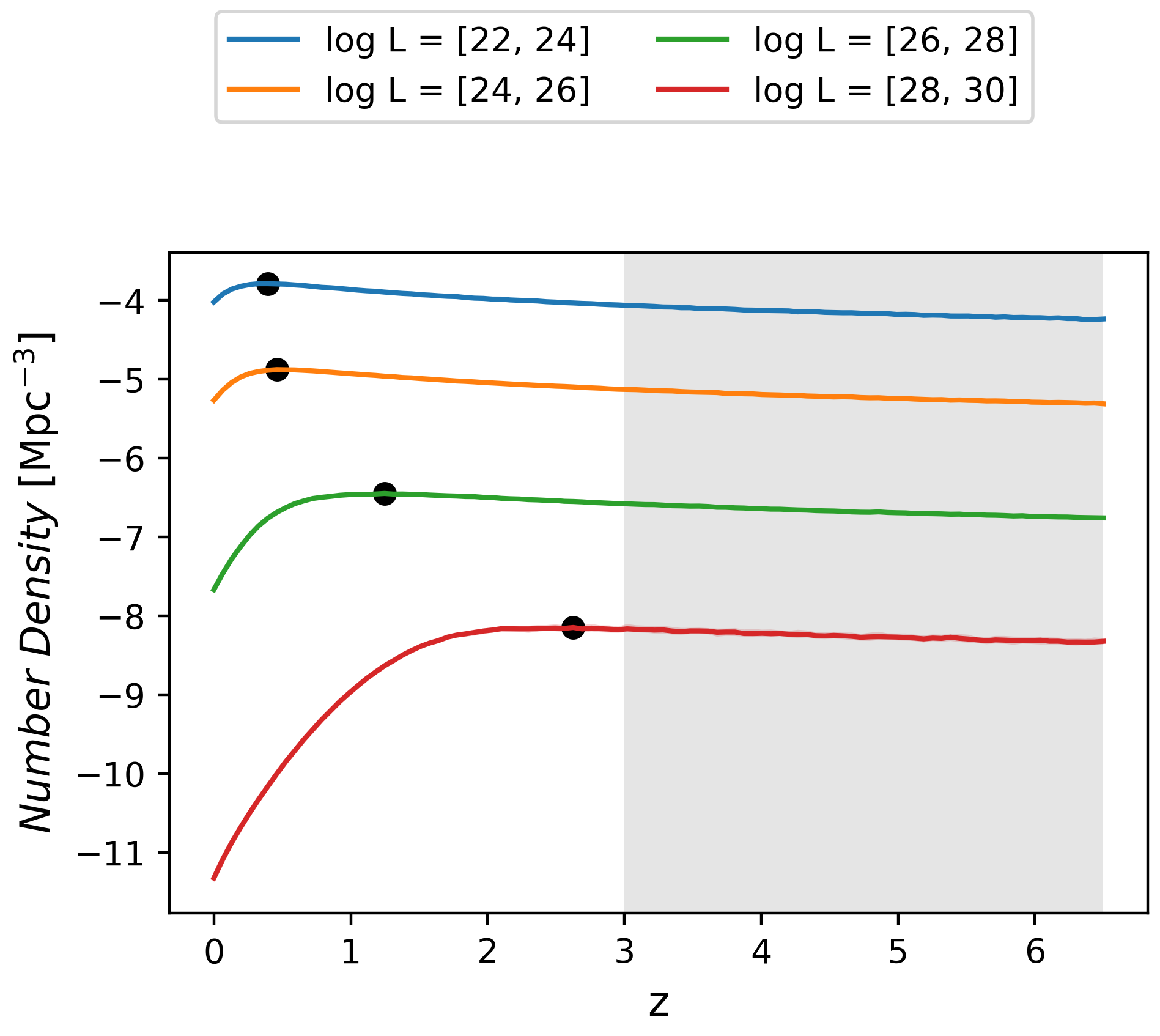}
\centering
\end{subfigure}
\hfill \newline
\begin{subfigure}{.5\textwidth}
\includegraphics[width=0.8\textwidth]{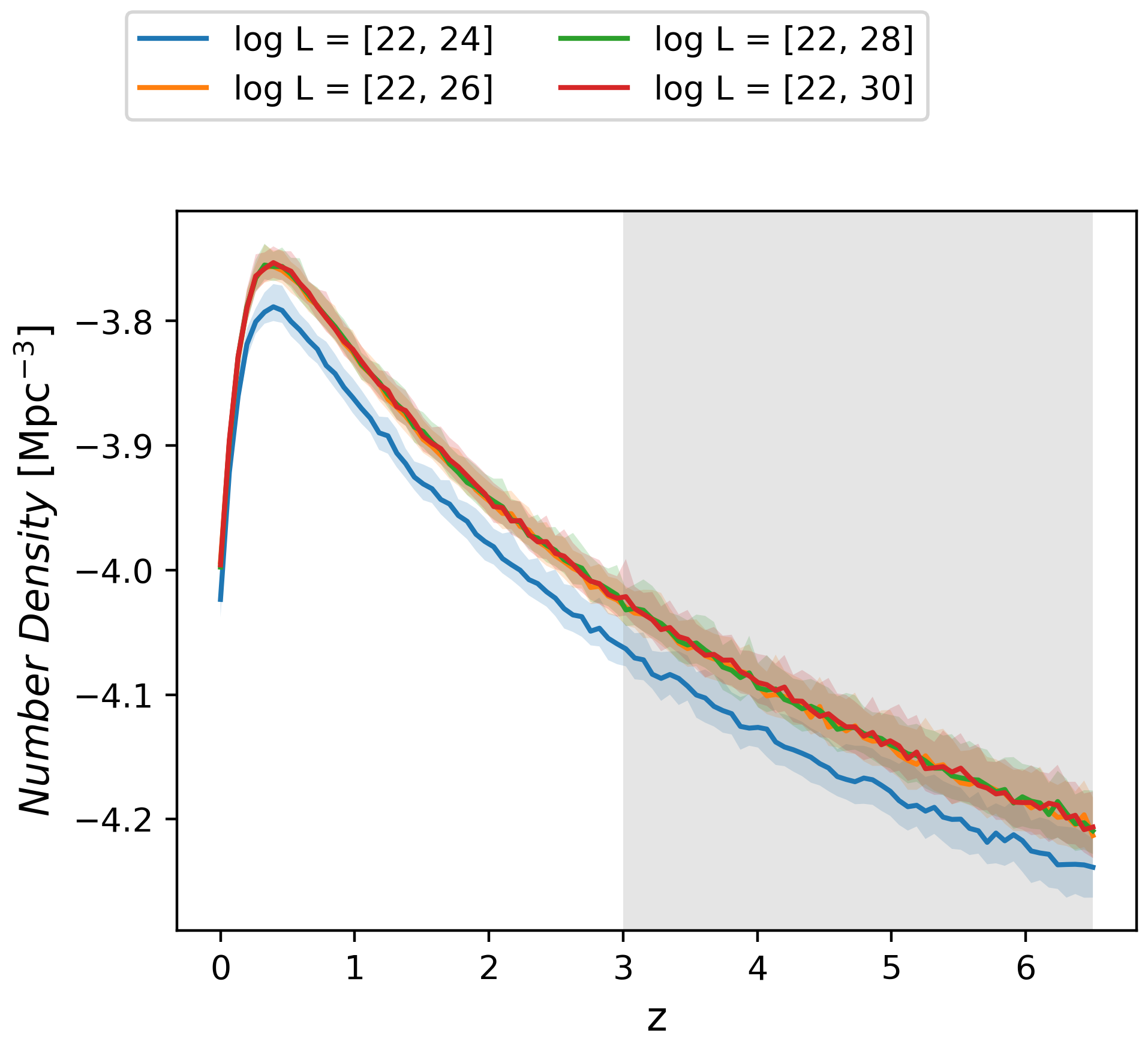}
\centering
\end{subfigure}
\begin{subfigure}{.5\textwidth}
\includegraphics[width=0.8\textwidth]{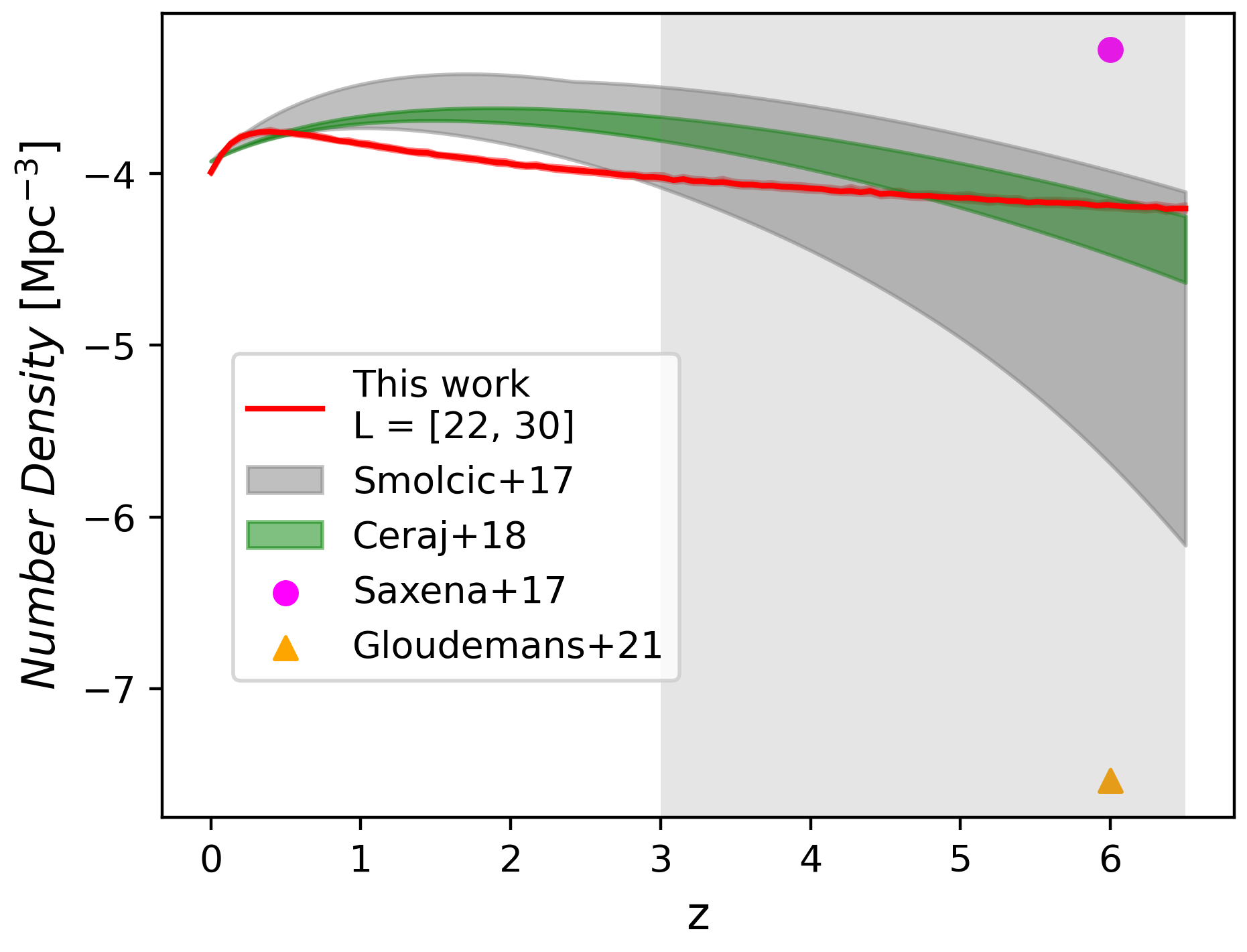}
\centering
\end{subfigure}
\caption{\textit{Upper panel:} Number density calculated at $1.4 \ \mathrm{GHz}$, for a set of different luminosity ranges  of same width, as denoted in the legend above the figure. The black dots represent the maximum value of each line. \textit{Middle panel:} Number density at $1.4 \ \mathrm{GHz}$ calculated for a set of progressively increasing luminosity ranges, as denoted in the legend above the figure. \textit{Lower panel:} Number density at $1.4 \ \mathrm{GHz}$ as a function of redshift for a set of different surveys, denoted in the legend. The data-points denote the high-redshift quasar surveys as described in the text. The uncertainties in this work are calculated from the resulting samples within the parametric Bayesian method as $90\%$ quantiles. The uncertainties of the literature values are determined as maximum uncertainties of the parameters as described in the text. The shaded area in the plots denote higher redshift where the LF models are less constrained. The high-redshift quasar density from \citet{Gloudemans2021} is a lower limit as the luminosity range of the LF used in the calculation was smaller.}
\label{fig:NumDen}
\end{figure}

\begin{figure}
\centering
\begin{subfigure}{.5\textwidth}
\includegraphics[width=0.8\textwidth]{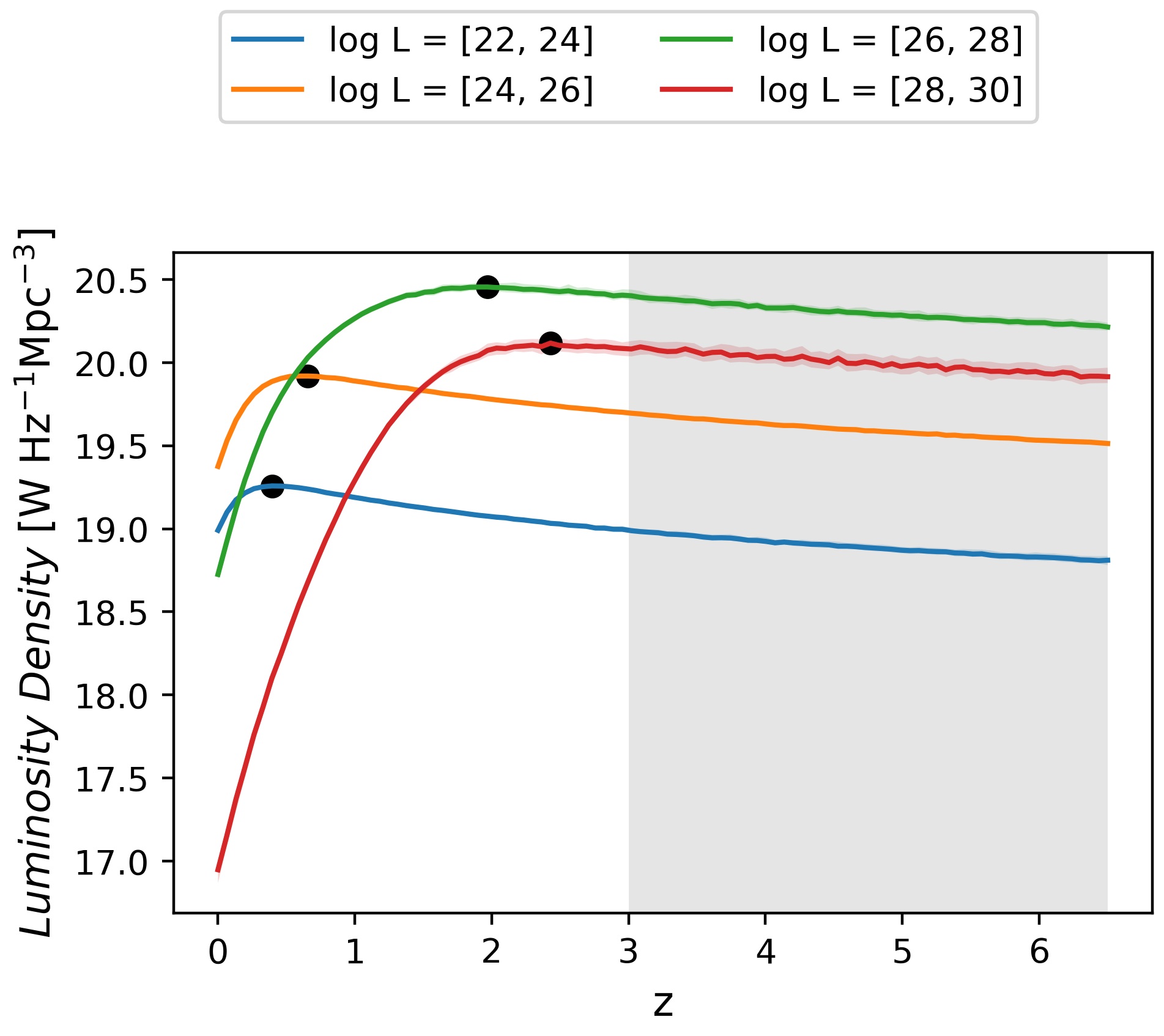}
\centering
\end{subfigure}
\hfill \newline
\begin{subfigure}{.5\textwidth}
\includegraphics[width=0.8\textwidth]{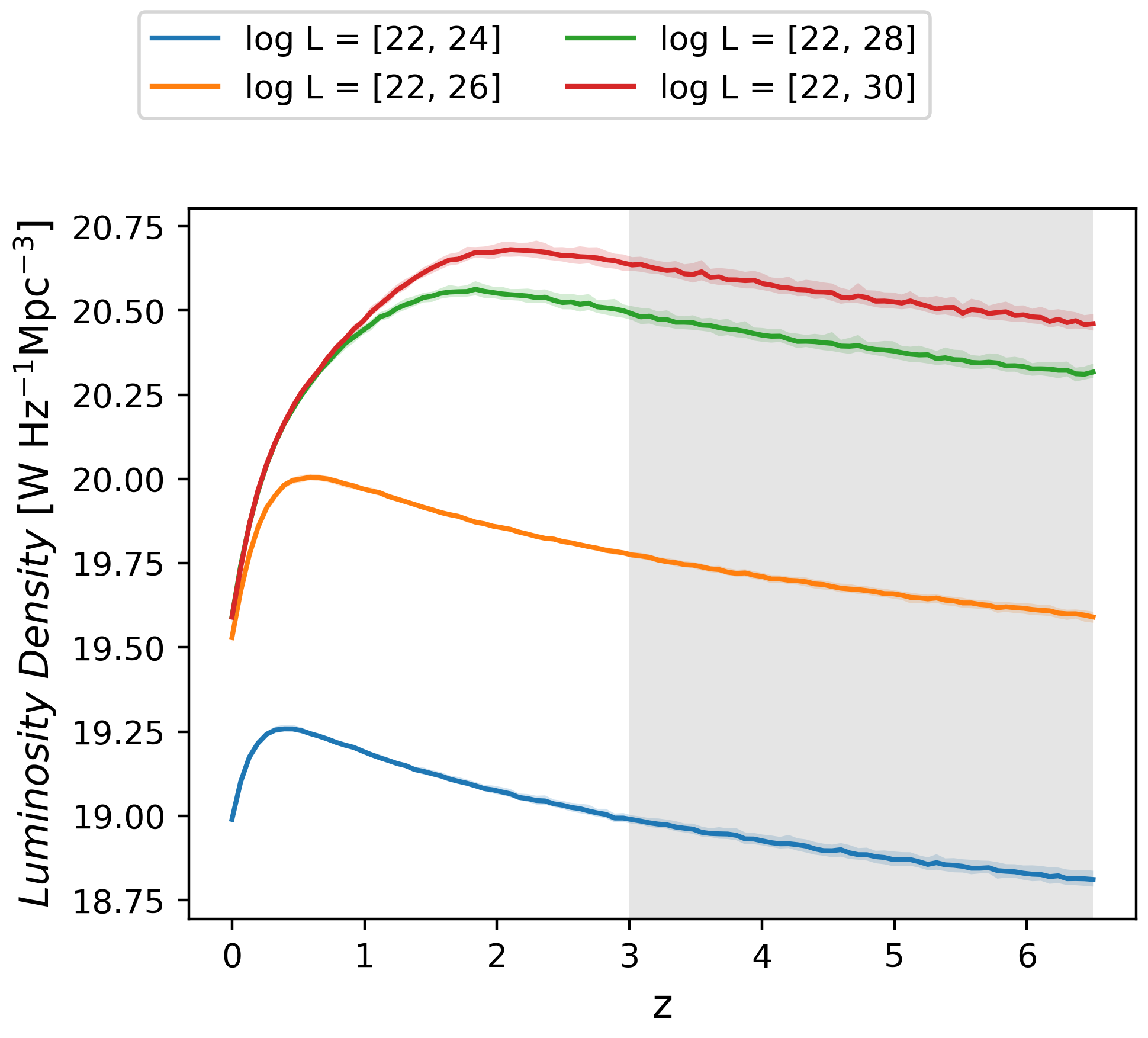}
\centering
\end{subfigure}
\begin{subfigure}{.5\textwidth}
\includegraphics[width=0.8\textwidth]{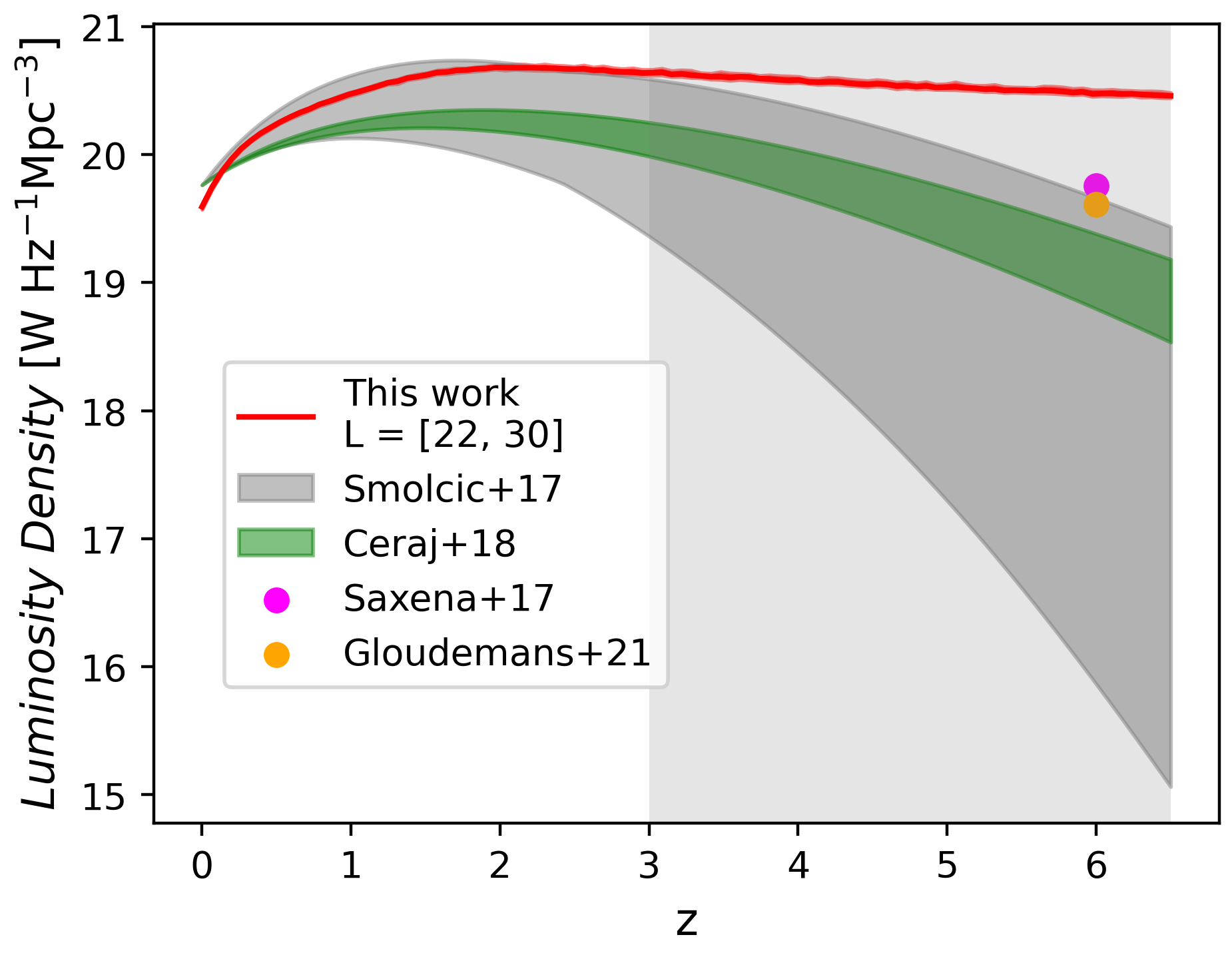}
\centering
\end{subfigure}
\caption{\textit{Upper panel:} Luminosity density at $1.4 \ \mathrm{GHz}$ calculated for a set of different luminosity ranges of same width, as denoted in the legend above the figure. The black dots represent the maximum value of each line. \textit{Middle panel:} Luminosity density at $1.4 \ \mathrm{GHz}$ calculated for a set of progressively increasing luminosity ranges, as denoted in the legend above the figure. \textit{Lower panel:} Luminosity density at $1.4 \ \mathrm{GHz}$ as a function of redshift for a set of different surveys, denoted in the legend. The data-points denote the high-redshift quasar surveys as described in the text. The uncertainties in the figure follow those in Fig \ref{fig:NumDen}. The shaded area in the plots denote higher redshift where the LF models are less constrained.}
\label{fig:LumDen}
\end{figure}

Using the best-fitting LDDE model, we estimated the number density and the luminosity density of sources as a function of redshift. The number density of sources was calculated as:
\begin{equation}
D_N(z) = \int_{L_{Min}}^{L_{Max}} \Phi(L,z) \ \der L
\label{eq:DenN}
\end{equation}
where the luminosity range was chosen as $\log [L_{Min}, L_{Max}] / ( \mathrm{W/Hz}) = [22,30]  $. The luminosity density was calculated within the same luminosity range as:
\begin{equation}
D_L(z) = \int_{L_{Min}}^{L_{Max}} L \cdot \Phi(L,z) \ \der L
\label{eq:DenL}
\end{equation}
The densities are shown in Figures \ref{fig:NumDen} and \ref{fig:LumDen}, extrapolated up to a redshift of $z=6$ in order to compare them with the high-redshift quasar studies. We use the estimation of the quasar luminosity function at $z=6$ from \citet{Gloudemans2021} and calculated the densities via relations (\ref{eq:DenN} $\&$ \ref{eq:DenL}). Their luminosity function was estimated by combining the properties of radio quasars at $z=2$ with the UV-luminosity function at $z=6$, assuming that the fraction of radio loud quasars remains constant from $z=2$ to $z=6$, as described in detail in the paper. Since the luminosity function from \citet{Gloudemans2021} spans a smaller luminosity range, the number density should be considered a lower limit. This effect however is not so important when calculating the luminosity density, as the integrated function is weighted by the value of luminosity. Another comparison was made with the high-redshift luminosity function of quasars from \citet{Saxena2017} predicted by the model developed within their work. The semi-analytical model uses black hole mass functions and Eddington ratio distribution, taking into account the energy losses due to synchrotron, adiabatic and inverse Compton processes, in order to predict the radio LF. The model also includes radio jets with powers determined via black hole mass and Eddington ratios. The radio LF was compared to observational data at $z = 2$ providing satisfactory results, and then extended to $z = 6$. The details of the model are described in detail in their work. The LFs resulting from the model were again integrated via relations (\ref{eq:DenN} $\&$ \ref{eq:DenL}) to obtain the densities.

We also show number and luminosity densities obtained from LF models of \citet{Ceraj2018} and \citet{Smolcic2017c}. Although, as described in these papers, the luminosity functions do not reach such high redshifts, we extrapolated them to $z=6$ in order to compare them with the high-redshift quasar surveys. The models assume that the evolution changes with redshift in analogy with the model by \citet{Novak2017} described in relation (\ref{eq:novakmodel}). The uncertainties plotted in the figure are the maximum between the $1\sigma$ uncertainties in parameters $\alpha$ and $\beta$. At lower redshifts the results are consistent, with differences arising at high redshifts (of $z \approx 5$). This is due to the difference in LF models used to describe the data. A slight difference in number density at low redshifts is the consequence of slightly different normalization of the local luminosity function. 

An interesting aspect of the luminosity density is the flattening at high redshifts. This effect is due to the bump present in the LDDE model at the high-luminosity end of the sample. To further illustrate this, in Figs \ref{fig:NumDen} and \ref{fig:LumDen}, we also plot the number and luminosity density using different luminosity ranges, each spanning progressively higher luminosities. Since the flattening occurs only when the upper luminosity boundary is high, we conclude that the high luminosity sources, responsible for the bump in the LF model, are responsible for the flattening of the luminosity density. The maximum values of these functions also change with luminosity bins. For luminosity density at $1.4 \ \mathrm{GHz}$, these values equal:
\begin{equation}
z = 0.39 \pm 0.05 \ , \ \ \ \log L \in [22,24]
\end{equation}
\begin{equation}
z = 0.60 \pm 0.03 \ , \ \ \ \log L \in [24,26]
\end{equation}
\begin{equation}
z = 1.98 \pm 0.2 \ , \ \ \ \log L \in [26,28]
\end{equation}
\begin{equation}
z = 2.51 \pm 0.06 \ , \ \ \ \log L \in [28,30]
\end{equation}
The values were estimated by taking the mean value from $5$ random samples of the posterior.

\subsection{Stellar mass dependent difference in evolution} \label{Sect:MassLFs}
\begin{figure*}
\centering
\includegraphics[width=0.95\textwidth]{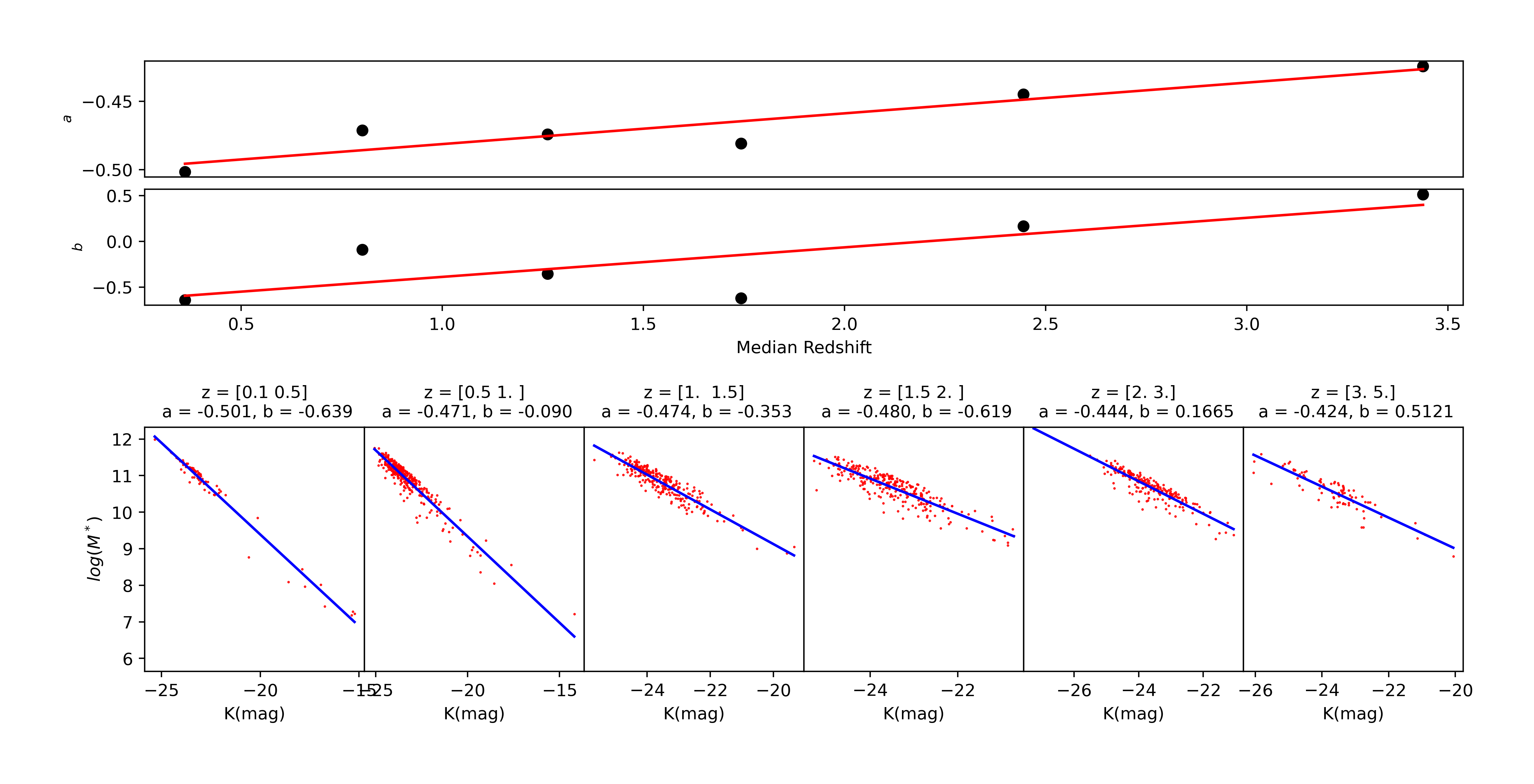}
\centering
\caption{The calibration of mass-to-light correlation between the absolute K-band magnitude and stellar mass obtained from the COSMOS2015 catalogue for the subset of AGN sources. The assumed functional form of the correlation is $M^* = a(z) K + b(z)$ as described in the text. \textit{Bottom:} The dependence of stellar mass $M*$ on K-band magnitude. The blue line shows the linear regression fit performed for each redshift bin independently. The range of each redshift bin is given above the corresponding plot, as well as the resulting correlation parameters. \textit{Top:} The resulting correlation parameters as a function of redshift. The red line shows the linear regression performed on these values in order to determine the redshift dependence of the parameters.}
\label{fig:StellarMass_calibration}
\end{figure*}

\begin{figure*}
\centering
\includegraphics[width=0.95\textwidth]{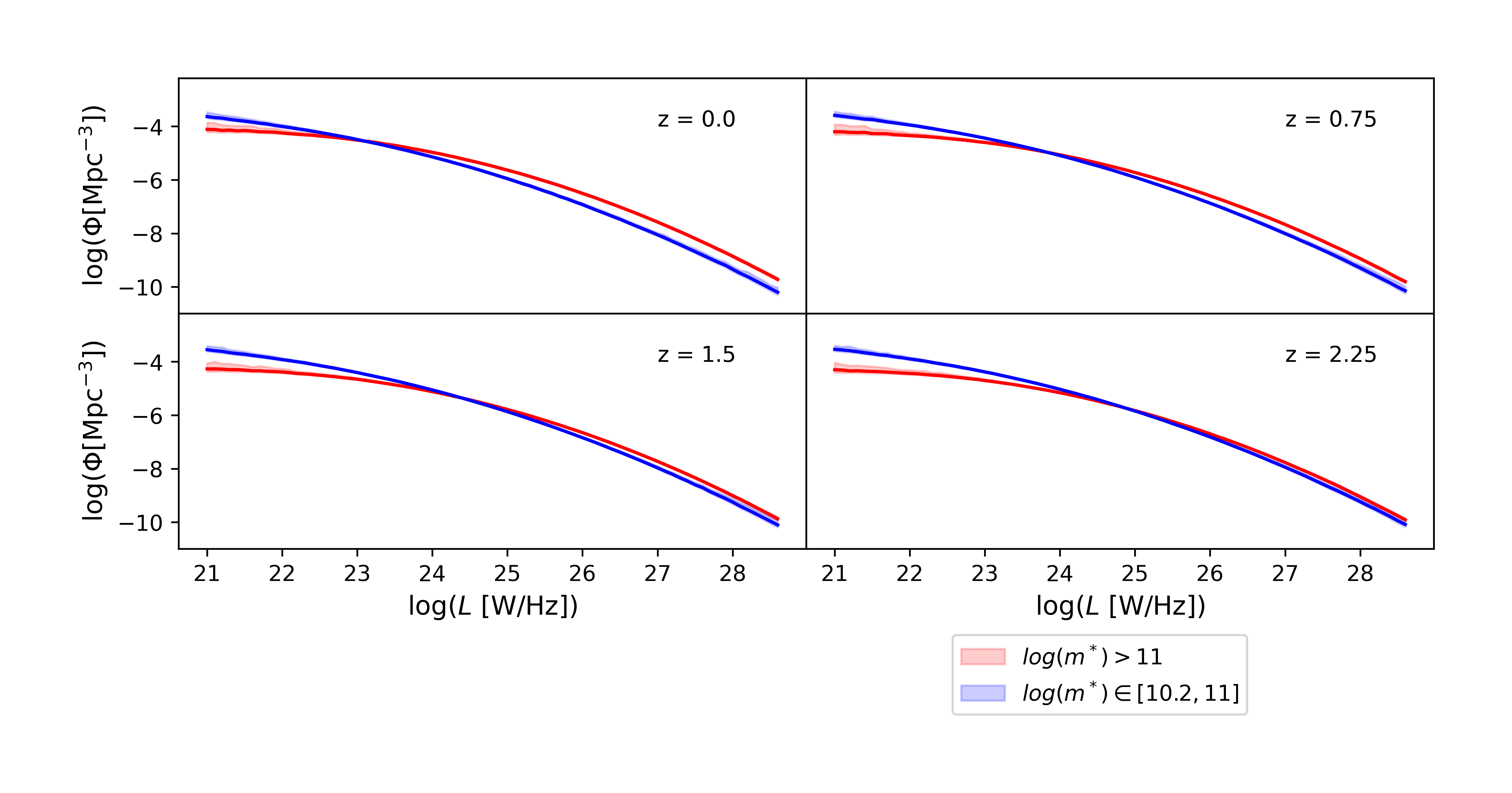}
\centering
\caption{Luminosity functions for the high and low mass sub-sample. The uncertainty plotted in the figure is the $68.2$ quantile. The model of evolution is the PDE model. The redshift of each subplot is given in the figure.  }
\label{fig:StellarMass_LFs}
\end{figure*}

In order to assess the dependence of the LF evolution on stellar mass, we divided our sample into two sub-populations of high and low mass galaxies. Since the XXL-N survey contained no stellar mass estimates, we excluded it from these considerations. Here we reasoned that the intermediate surveys are already constrained by the XXL-S field, so this simplification is not crucial. The stellar mass estimates for the COSMOS field come from the COSMOS2015 catalogue (\citealt{Laigle2016}) and are calculated from spectra as described in \citet{Laigle2016}. The XXL-S field stellar mass estimates are determined by SED fitting as described in XXL Paper XXXI. The fields from \citealt{Willott2001} lacked stellar mass estimates but contained apparent K-band magnitudes, via a publicly available catalogue\footnote{https://astroherzberg.org/people/chris-willott/research/}. The complete publicly available catalogue, for all three shallow fields $7C$, $6CE$ and $3CRR$, contained $181$ sources. The $7C$ had complete K-band magnitudes at $z > 1.2$, so a threshold was imposed on this data set, and a correction made during the calculation of likelihood. The $3CRR$ field had $69/96$ sources with K-band magnitudes at $z > 0.05$. This incompleteness was incorporated via a correction function to the likelihood function. The $6CE$ field had complete K-band magnitude data. The details on the catalogues are found in \citet{Willott2003K}. In order to estimate the stellar masses we used the relationship between the stellar mass of the galaxy and its K-band magnitude reported in the literature (e.g. \citealt{Arnouts2007}). Since we are dealing with a sub-population of purely AGNs, we re-calibrated the stellar mass to K-band correlation. For this purpose, we used the COSMOS2015 catalogue that contains both these values. We took a subset of the catalogue containing purely AGNs, based on radio excess, as previously described in Sect. \ref{sect:AGNSample}, and re-plotted the dependence of stellar mass on the absolute K-band magnitude, as shown in Fig \ref{fig:StellarMass_calibration}. Following \citet{Arnouts2007} we allowed for a redshift dependent correlation via two free redshift-dependent parameters as: $\log (M^*) = a(z) K + b(z)$. First we performed a linear regression fit on every redshift subset independently. The resulting parameters $a$ and $b$ differed across redshift bins. By performing another linear regression on these values we assessed the redshift dependence of the parameters. The resulting correlation parameters thus equalled\footnote{The catalogue of stellar mass values used in this work is available as a table uploaded to the CDS. The columns of the catalogue are described in the appendix. This catalogue is a compilation from other surveys, except for the stellar masses for the 3CRR, 7C and 6CE surveys which were calculated within this work.}:
\begin{equation}
a(z) = 0.0224 \cdot z  -0.503
\label{eq:MassCala}
\end{equation}
\begin{equation}
b(z) = 0.3226 \cdot z  -0.711
\label{eq:MassCalb}
\end{equation}
Furthermore to eliminate any systematic error arising from the incompleteness of our sample due to a finite detection limit, we removed the lowest-mass galaxies from our sample. The two stellar mass sub-populations therefore spanned mass intervals of $\log(M^*) \in [10.2, 11]$ and $\log(M^*) > 11$, ranging in log-luminosities, for both subsets, from $\approx 22$ to $\approx 28$. The difference in evolution can be seen in Fig. \ref{fig:StellarMass_LFs}. We modelled the evolution as a simple PDE evolution since we were interested only in tracing the difference between the sub-populations. The difference in evolution exists and is larger than the $68.2$ quantiles plotted in the figure. The parameter of PDE evolution (see equation \ref{Evol:PDEPLE}) equalled $\alpha_D = 0.23 ^{+ 0.13} _{- 0.13}$ for the low-mass sample and $\alpha_D = -0.38 ^{+ 0.11} _{- 0.12}$ for the high-mass, but the differences of the functions arise also as a result of the complete LF shape. As a final precaution, we repeated the LF fitting without the $7C$ survey, since this survey had the largest incompleteness. The results remained qualitatively the same. All in all, the difference in LFs could point towards some kind of bimodality within our AGN sample which is a function of host galaxy stellar mass. The details of this bimodality, however, need to be investigated further.

\subsection{Source counts}
\label{Sect:SC}
\begin{figure*}
\centering
\includegraphics[width=0.55\textwidth]{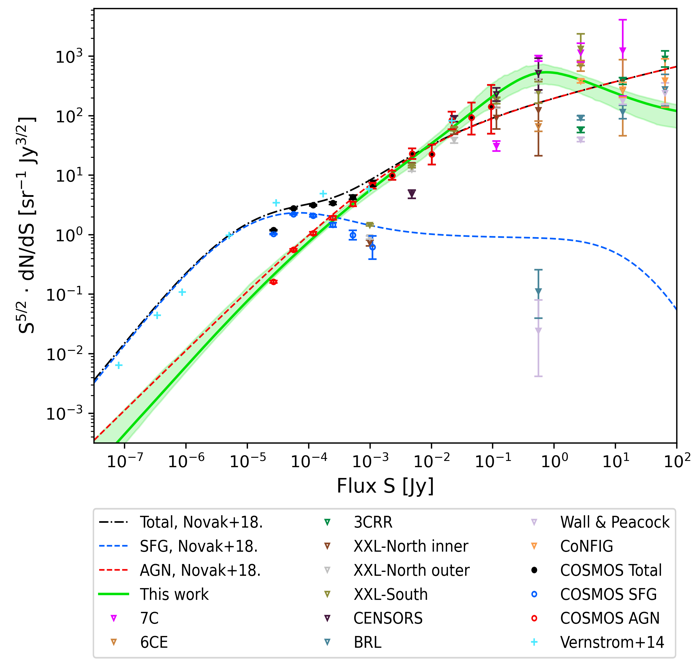}
\caption{The source counts model together with data points obtained directly from the catalogues. The green dashed line denotes the model obtained from LFs constrained within this work. The errors were determined by selecting $500$ samples from the posterior. The red, blue and black lines denote models from \citealt{Novak2018} obtained from LFs for AGN, SFG and the total population respectively. Data-points represent the source counts obtained from the catalogues as denoted in the legend. All the catalogues are the same as described in Sect. \ref{sec:Data} except the ones denoted as Vernstrom+14, which were taken from another study by \citealt{Vernstrom2014}. COSMOS SFG are sources from the COSMOS catalogue not selected by the radio excess threshold described in Sect. \ref{sec:Data}. The outlier data points are the effect of finite detection limit.}
\label{fig:SC}
\centering
\end{figure*}

Using the shape of the modelled LFs, it is possible to construct the AGN source counts of our sample. We show in Fig. \ref{fig:SC} the AGN source counts obtained from our model of the LF, by drawing $500$ parameter samples from the posterior. From the definition of the luminosity function $\Phi(L,z)$, the number of sources $\Delta N$ in each flux bin at a certain value of redshift was obtained as:
\begin{equation}
\label{eq:SC1}
\Delta N = \Phi (L,z) \frac{dV}{dz} \Delta \log L \: dz,
\end{equation} 
where $dV/dz$ is the differential comoving volume, $\Delta \log L$ luminosity decade and $dz$ the redshift bin. The number of sources obtained in each flux bin was summed over all redshift bins and then normalized with counts expected in a static Euclidean Universe. For comparison, we also show source counts from an earlier study by \citet{Vernstrom2014} and the model obtained from the LF evolutionary model by \cite{Novak2018}, as this is the model constrained by the deeper COSMOS survey, and as such constrains the low luminosity end of the sample best. The source counts by \citet{Vernstrom2014} were obtained from $3 \ \mathrm{GHz}$ data observed by Karl G. Jansky Very Large Array, directed towards the Lockman Hole. The source counts were constructed using the method of probability of deflection to reach deeper values of flux, as described in the paper. The model by \citealt{Novak2018} was constructed from the LFs that have pure luminosity evolution, with different parameters, for star forming galaxies (SFG) and the AGN population as described in detail in the paper.

\section{Discussion}
\label{sect:Discuss}
We have modeled the luminosity functions via Bayesian parametric method, using a composite sample consisting of multiple surveys of varying area and depth ($z < 3$ and $\log L \in [22,29])$, which together span a large interval in both redshift and luminosity. We compared a set of LF models and concluded that by all used criteria, the LDDE model was the preferred one. This result is broadly consistent with earlier studies, where the difference in evolution is observed between sub-samples of high and low luminosity.

\subsection{Evolution of AGN sub-populations in the literature}

The difference in the evolution of the high and low luminosity end of the sample is reported throughout the literature (\citealt{Hook1998}, \citealt{Waddington2001}, \citealt{Willott2001}, \citealt{Clewley2004}, \citealt{Sadler2007}, \citealt{Smolcic2009}, \citealt{Donoso2009}, \citealt{Padovani2017}). There are however differences in the adopted LF models. As already stated, \citet{Willott2001} use a bimodal model where the shape and evolution of the high and low luminosity end have a different functional form. \citet{Smolcic2009} model only the low-end of the AGN sample using a superposition of luminosity and density evolution, analogous to relation \ref{Evol:PDEPLE}, showing a modest evolution of this sample compared to the high-luminosity studies. Furthermore, studies are often performed via non-parametric methods (e.g. \citealt{Waddington2001}, \citealt{Sadler2007}, \citealt{Donoso2009}, \citealt{Rigby2015}) so no functional form is assumed for the LF. Even so, there is a difference in evolution which seems to be a function of luminosity, which makes these results consistent with our work.

A difference in evolution between two subsets of AGN was observed in a study by \citet{Pracy2016} using the Faint Images of the Radio Sky at Twenty-cm survey matched with the Sloan Digital Sky Survey. The complete AGN sample, was divided into HERGs and LERGs based on optical spectra. There was an observed difference in evolution in the double power-law LF assuming alternatively both PDE or PLE evolution, where the LERG population evolved slowly as opposed to the more rapid evolution of the HERG subsample. Although the comparison is not exact, due to a difference in classification, this difference in evolution is consistent with our results where the evolution depends on luminosity.

\citet{Padovani2015} divided their Extended Chandra Deep Field-South Very Large Array sample into RQ and RL AGN based on relative strength of radio emission at $1.4. \ \mathrm{GHz}$ as described in the text, where RL sources correspond to the ones with radio excess\footnote{The classification of RL and RQ sources is not consistent throughout the literature. See \citet{Padovani2017}, Sect. 2.1.3. for details.}. RL AGNs correspond mostly to jet-mode AGNs, and RQ to radiative mode. A difference in evolution between these two sub-populations was observed, where the RL sample exhibited a peak at $z = 0.5$ after which their numbers declined as opposed to the RQ sample. These findings are also consistent with this work.

A study by \citet{Ocran2021} of the ELAIS N1 field observed with the GMRT at $610 \ \mathrm{MHz}$ divided the complete sample into RQ and RL AGN, based on a combination of multi-wavelength criteria as described in their text. The evolution was modeled as PLE for the sub-samples and a difference in evolution was observed, where RL AGN evolved more strongly. This is again consistent with our results.

Similar conclusions concerning AGN evolution are also obtained within X-ray astronomy. An example is the study by XXL Paper VI, using a composite set of fields: MAXI, HBSS, XMM-COSMOS, Lockman Hole, XMM-CDFS, AEGIS-XD, Chandra-COSMOS, and Chandra-CDFS. The LFs were modeled using an AGN sample observed within the X-ray part of the spectrum in the $5-10 \ \mathrm{keV}$ band. The model comparison was done also within the Bayesian framework, comparing AIC and BIC, resulting in LDDE being the best-fitting model.

On the other hand, some of the studies, such as \citet{Yuan2016} argue via theoretical arguments against the LDDE model, explaining the phenomenology within the framework of luminosity and density evolution mixture model. This is not consistent with our results.

\subsection{The existence of two AGN sub-populations}

A trend throughout the literature is the separation of the full AGN sample into sub-samples, be it RL and RQ, based on the relative strength of the radio emission (e.g., \citealt{Padovani2015}), high HERG and LERG, based on optical spectral lines (Paper XXXVI), flat and steep spectrum sources, based on the spectral index (\citealt{Wall1975}, \citealt{Massardi2010}, \citealt{Bonato2017}), or any other criterion. Although the LDDE model, preferred in this work, assumes a continuous change in evolution with regards to luminosity, this does not exclude the existence of two sub-populations. Firstly, the strength of the model selection criteria between the model from \citet{Willott2001} and LDDE is not as strong compared to the simple PDE and LDE models, and the difference could be a consequence of the larger number of parameters of the Willott model. More importantly, if the sub-populations are not selected with a simple luminosity threshold, different fraction of each population can be present at different luminosities. This could lead to the observed continuous change in evolution with luminosity, present in the LDDE model. Concentrating on the underlying physical processes, these results are therefore still consistent with the picture outlined in the introduction, where there exist two distinct modes of accretion: the radiatively efficient mode, and the radiatively inefficient mode (\citealt{Hardcastle2007}, \citealt{Heckman_Best2014}, \citealt{Narayan1998}, \citealt{Shakura1973}). Although the analogies are not exact, the radiatively efficient mode would correspond to the high luminosity end or the HERG sub-sample, while the inefficient mode to the low luminosity end or the LERG sub-sample.

\subsection{Kinetic luminosity}

\begin{figure}
\centering
\includegraphics[width=0.45\textwidth]{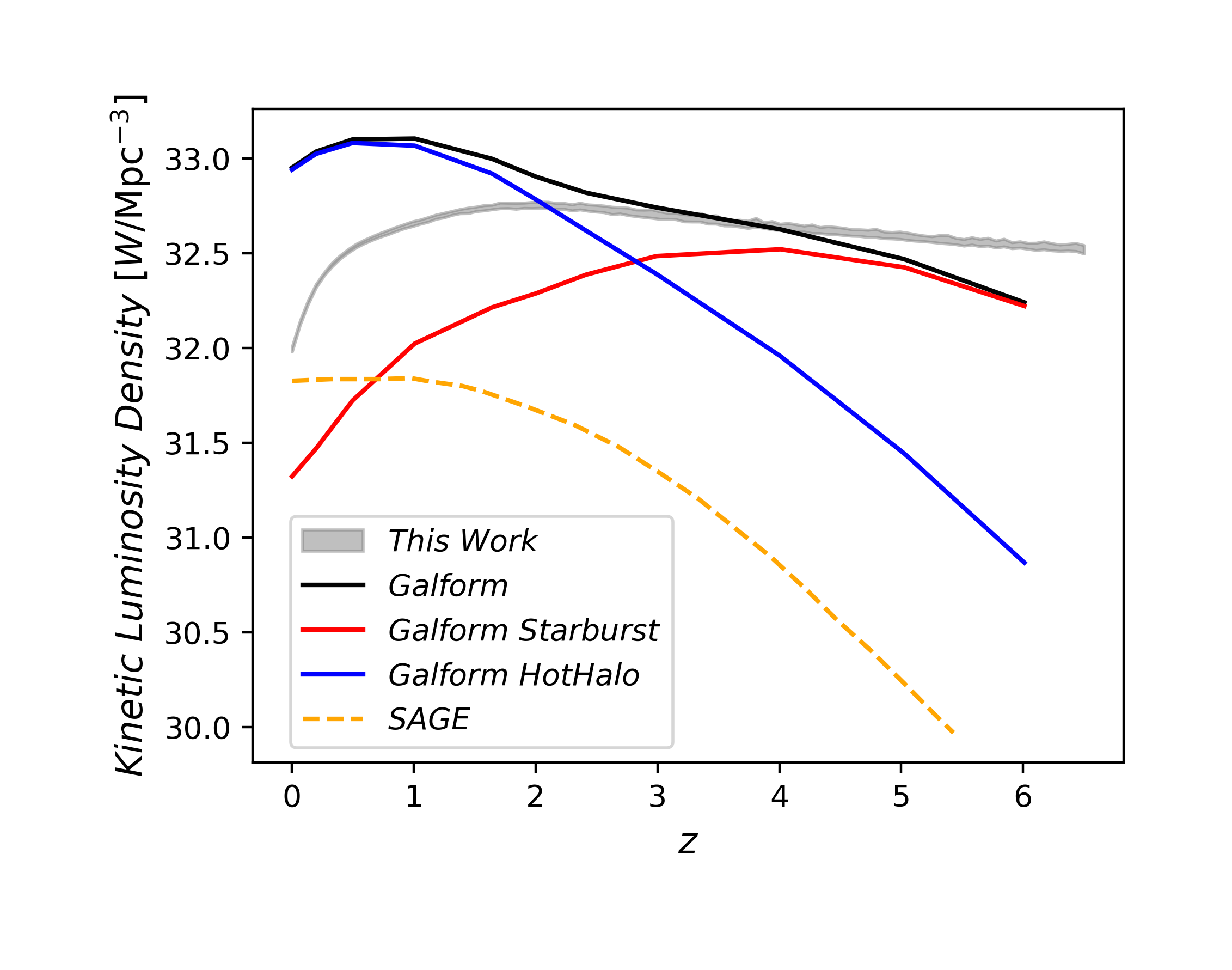}
\centering
\caption{Kinetic luminosity density as a function of redshift given in grey. The uncertainties are calculated from the resulting samples of the parametric method as $90\%$ quantiles. Black red and blue lines correspond to the predictions from GALFORM. The black line is the total density, while the red and blue lines denote the hot-halo and starburst modes, respectively. The orange dashed line represents the SAGE model.}
\label{fig:KinDen}
\end{figure}

Apart from the observed radio emission, a large part of the energy stored in the AGN jets is given to the environment kinetically via work performed by jet expansion (e.g. \citealt{Smolcic2017c}). In order to assess this power and to gain insight into how the feedback of AGNs evolves through cosmic time we investigated the kinetic luminosity of our sample. Using the correlation from \citet{Ceraj2018} (see also \citealt{Smolcic2017c}) we determined the kinetic luminosity from the radio luminosity following the relation:
\begin{equation}
\log(L_{Kin}) = 0.86 \cdot \log(L_{1400 \ \mathrm{MHz}}) + 14.8 + 1.5 \cdot \log (f)
\label{eq:KinLum}
\end{equation}
where $f$ was introduced by \citet{Willott1999} in order to incorporate all the possible systematic errors, and was determined to be in the range $1-20$. Following \citet{Ceraj2018}, we set it to $15$.  It should be noted however that the parameter changes the luminosity by a multiplicative constant factor. The corresponding kinetic luminosity density will therefore shift systematically on the y-axis but the shape of the redshift dependence will remain the same. The uncertainties are also large enough to include the star-forming component of radio emission. We calculated the kinetic luminosity density as a function of redshift:
\begin{equation}
D_{kin}(z) = \int_{L_{Min}}^{L_{Max}} L_{kin} \cdot \Phi(L,z) \ \der L
\label{eq:DenKin}
\end{equation}
Here we again used the samples from the LDDE model as described in the last subsection. The resulting plot is shown in Fig. \ref{fig:KinDen}. Apart from our observational results, we also show the estimated kinetic luminosity density obtained from the GALFORM model (\citealt{Fanidakis2012}) and the SAGE model (\citealt{Croton2006}). The GALFORM model assumes two different modes of black hole accretion and subsequently two different evolution modes through cosmic time. The first mode is the starburst mode where accretion arises from galaxy mergers or instabilities within the disk. The second mode is the hot-halo mode accreting matter from the hot halo onto the central black hole. An interesting aspect of the GALFORM model is the flattening between the observed kinetic luminosity and the total and starburst modes of the GALFORM model at redshifts $z \approx 3-4$, not present in the SAGE model. The SAGE model, which includes the feedback mechanism, has black hole accretion rate $\dot{m}$ as one of its results. Following \citet{Croton2016} and \citet{Ceraj2018}, we calculated the kinetic luminosity from this value as: $0.1 \cdot \dot{m} \cdot c^2$ multiplying this by $0.08$ which was the radio mode efficiency parameter. The factor $0.1$ is the standard value found in the literature, falling between the efficiency expected for a non-spinning and maximally spinning black hole (\citealt{Croton2016}). Our comparison with the SAGE model gives non-consistent results. Even if we ignore the absolute values of the functional forms, which can be explained with the uncertainty factor $f$ given in relation (\ref{eq:KinLum}), the shape (i.e. redshift dependence) is different between the model and observations. Furthermore, Fig. \ref{fig:KinDen} shows that the two models give different kinetic luminosity estimates. The differences between models, and between our observational results, are probably due to the assumptions made in the models.

\subsection{Downsizing and Feedback}

The described trend of different evolution and cutoffs can be explained via cosmic downsizing, where the more massive black holes form earlier than the less massive ones (e.g., \citealt{Rigby2015}). This trend is, at first glance, not consistent with the hierarchical model, where larger black holes are the product of merging, but the apparent inconsistency can be explained by a switch in the mode of accretion between the efficient cold gas accretion to inefficient hot gas accretion via process called feedback, where inefficient accretion starts to dominate at low redshifts (e.g., \citealt{Heckman_Best2014}, \citealt{Rigby2015}). In other words, the accretion onto the central black hole is a complex process, where the rate of accretion can lead to a feedback effect, slowing it down or even quenching it. Consequently, there is a switch between the high efficiency accretion to low efficiency accretion. This, in turn, affects the properties of AGN through cosmic time, and hence the AGN evolution. Since in this work we argue for a luminosity dependent evolution of AGN, it is consistent with a physical picture requiring feedback. In short, difference in evolution as a function of AGN luminosity shows that the physics of AGN evolution depends on the accretion rate. This is consistent with the picture of AGN accretion which incorporates feedback. This also places our results in line with other publications where feedback was either deduced indirectly via scaling relations of the host galaxy and its black hole (\citealt{Magorrian1998}, \citealt{Ferrarese_Merritt2000}, \citealt{Gebhardt2000}, \citealt{Graham2011},  \citealt{Sani2011}, \citealt{Beifiori2012},  \citealt{McConnell_Ma2013}), or observed more directly via galactic winds (e.g., \citealt{Nesvadba2008}, \citealt{Feruglio2010}, \citealt{Veilleux2013}, \citealt{Tombesi2015}) or X-ray cavities in galactic groups and clusters (\citealt{Clarke1997}, \citealt{Rafferty2006}, \citealt{McNamaraNulsen2007}, \citealt{Fabian2012}, \citealt{Nawaz2014}, \citealt{Kolokythas2015}). Lastly, the need for AGN feedback is also supported by simulations (\citealt{Fanidakis2012}, \citealt{Hirschmann2012}, \citealt{Croton2016}, \citealt{Harrison2018}).

\section{Summary and conclusion}
\label{sec:Sum}

We modelled the radio luminosity functions of AGNs using a composite survey of varying area and depth, namely the COSMOS, XXL-North, XXL-South, $7$C, $6$CE and $3$CRR fields, consisting all together of $4,655$ sources. This allowed us to constrain the luminosity functions both at high redshifts (up to $z \approx 3$) and at high luminosities ($\log L \in [22,29])$. The functions were modelled with emphasis on the parametric method within the Bayesian framework, which allowed us to select the best fitting model from a set of different shapes and evolutions. The best fitting model according to marginal likelihood comparison, as well as the AIC and BIC methods, was the LDDE model, Using the Jeffreys interpretation, evidence ratios varied from "strong" ($>10$) to "decisive" ($>100$). The parameter posteriors were determined from the Bayesian model fitting and the resulting values detrmined as listed in Table \ref{Tab:ParameterPosteriors}. The dependence of shape and evolution of the LFs on luminosity assumed by this model was discussed in its implications on the physical picture of AGN evolution through cosmic time. Although the change in evolution as a function of luminosity is continuous, this does not exclude the possibility of AGN sub-populations as different fractions of each sub-population can be found at different luminosities. We discussed the number density and luminosity density as a function of redshift. The shape of the best fitting LDDE model resulted in a flattening at higher redshifts that is not present in simpler models with pure density or luminosity evolution. We compared these results with high-redshift quasar surveys and found broad consistency. We calculated the kinetic luminosity density and compared it to model-estimated values finding some consistency with the GALFORM simulation, but not with the SAGE model. Furthermore, in order to assess the dependence of stellar-mass of host galaxies on AGN evolution, we divided our sample into subsets of different stellar mass and modelled the evolution using a simpler PDE model. The difference in LFs was observed that was larger than $65\%$ quantiles estimated from posterior samples. Taken together, all these results point to a physical picture of AGN evolution where a simple density evolution, luminosity evolution or a superposition of both is not enough to trace the details of AGN evolution. More complex models, either consisting of AGN sub-populations, or including a dependence on AGN luminosity, are needed.

\begin{acknowledgements}
We thank the anonymous referee for insightful comments that improved the quality of the paper. VS acknowledges the European Union's Seventh Framework programme under grant agreement 337595 (CoSMass). BS and VS acknowledge the financial support by the Croatian Science Foundation for project IP-2018-01-2889 (LowFreqCRO). ZI acknowledges support by the U.S. Fulbright Scholar Program and hospitality at the Ruđer Bošković Institute, Zagreb, Croatia. XXL is an international project based around an XMM Very Large Programme surveying two $25 \ \mathrm{deg}^2$ extragalactic fields at a depth of $\sim 6 \cdot 10^{-15}\ \mathrm{erg}\ \mathrm{cm}^{-2} \mathrm{s}^{-1}$ in the $[0.5-2]\ \mathrm{keV}$ band for point-like sources. The XXL website is \url{http://irfu.cea.fr/xxl}. Multi-band information and spectroscopic follow-up of the X-ray sources are obtained through a number of survey programs, summarized at \url{http://xxlmultiwave.pbworks.com/}. MP acknowledges long-term support from the Centre National d'Etudes Spatiales (CNES).
\end{acknowledgements}

\appendix
\counterwithin{figure}{section}

\section{Stellar mass catalogue description}
The columns of the catalogue of stellar masses, introduced in  Sect. \ref{Sect:MassLFs} (see footnote 2, available at. ...), submitted via CDS, are as follows:
\begin{itemize}
   \item Name: Name of the radio source
   \item z: Best available redshift for the source
   \item $S\_1400\_MHz$: Source flux at $1400 \ \mathrm{MHz}$ in $\mathrm{mJy}$
   \item Alpha: Spectral index of the source. Set to mean of respective field when not available
   \item Mstar: Stellar mass of the source. Determined as described in Sect. \ref{Sect:MassLFs}
   \item Survey: Name of original survey from where the source was taken. "C" denotes COSMOS, "XXL-N" and "XXL-S" denote the North and South XXL fields, "$3$" denotes the 3CRR field, "$6$" the 6CE and "$7$" 7C fields.  
\end{itemize}
This catalogue is a compilation from other surveys, except for the stellar masses of 3CRR, 7C and 6CE surveys. These were determined from magnitudes within this work, as described in the text.

\section{Non-optimal model fits}
\begin{figure*}
\centering
\includegraphics[width=0.95\textwidth]{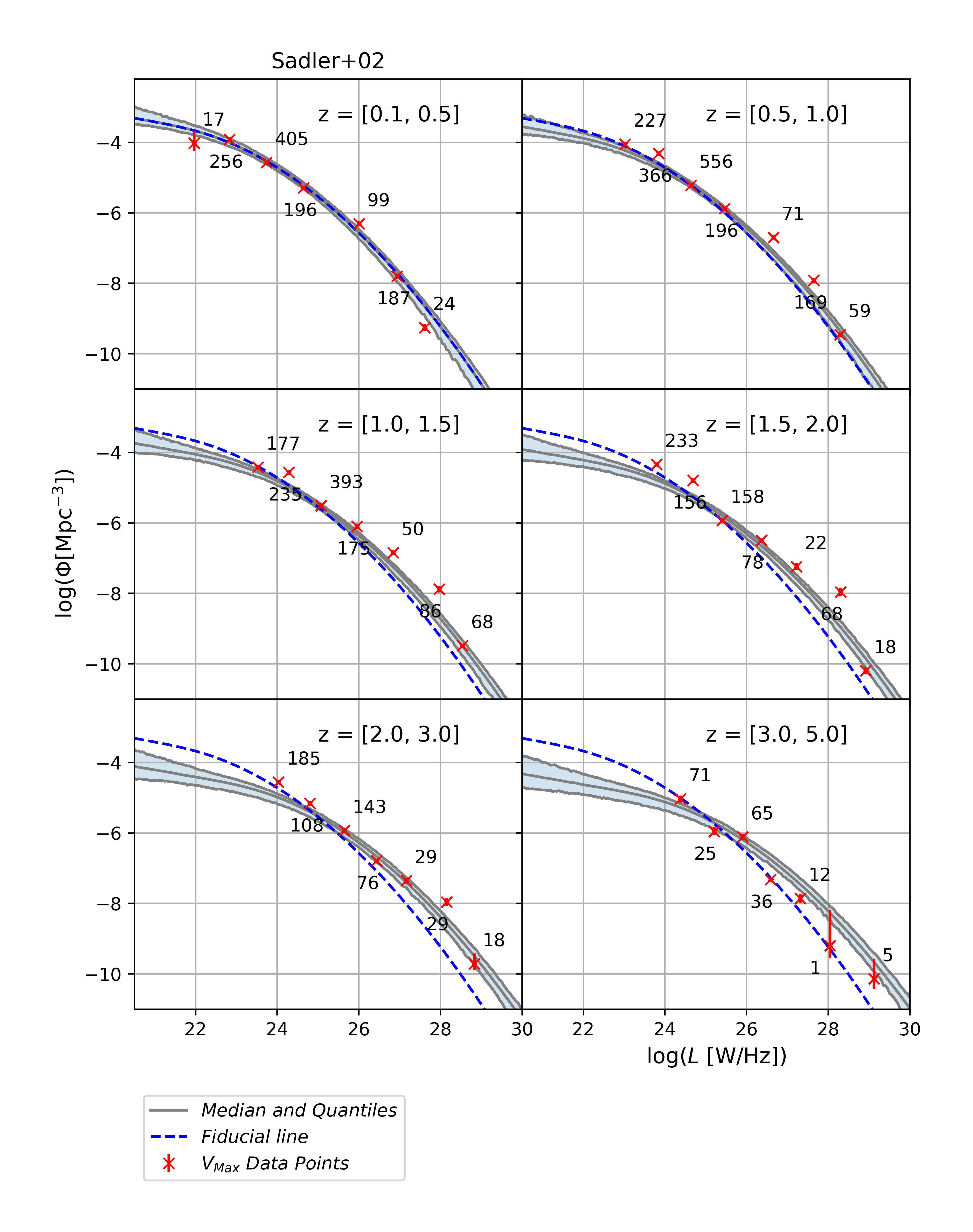}
\centering
\caption{The Sadler$+02$ model LF fit. The model evolution is described in relation (\ref{Evol:PDEPLE}). The notation follows Fig. \ref{fig:LFs_LDDE}.}
\label{fig:LFs_App1}
\end{figure*}

\begin{figure*}
\centering
\includegraphics[width=0.95\textwidth]{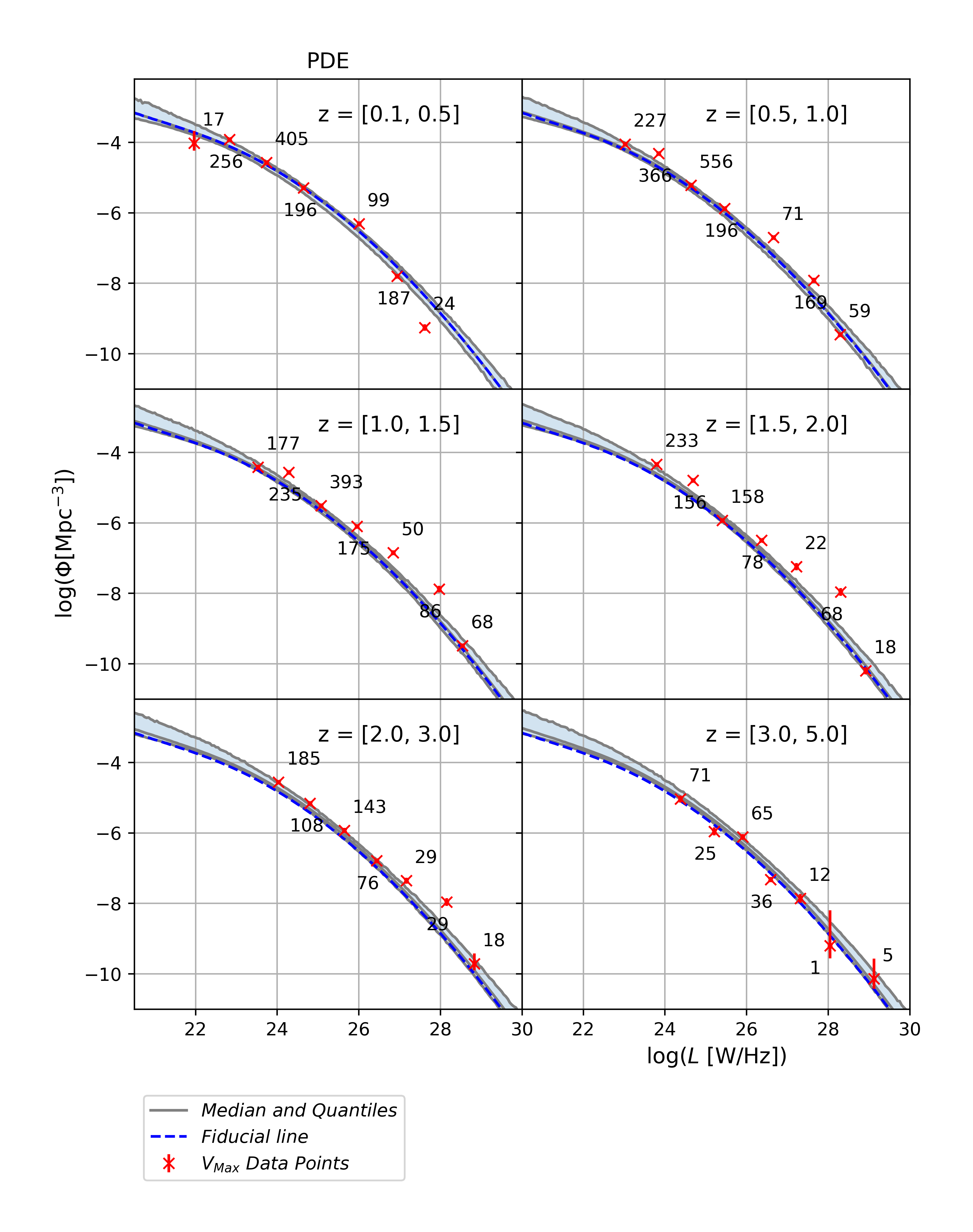}
\centering
\caption{The PDE model LF fit. The notation follows Fig. \ref{fig:LFs_LDDE}.}
\label{fig:LFs_App2}
\end{figure*}

\begin{figure*}
\centering
\includegraphics[width=0.95\textwidth]{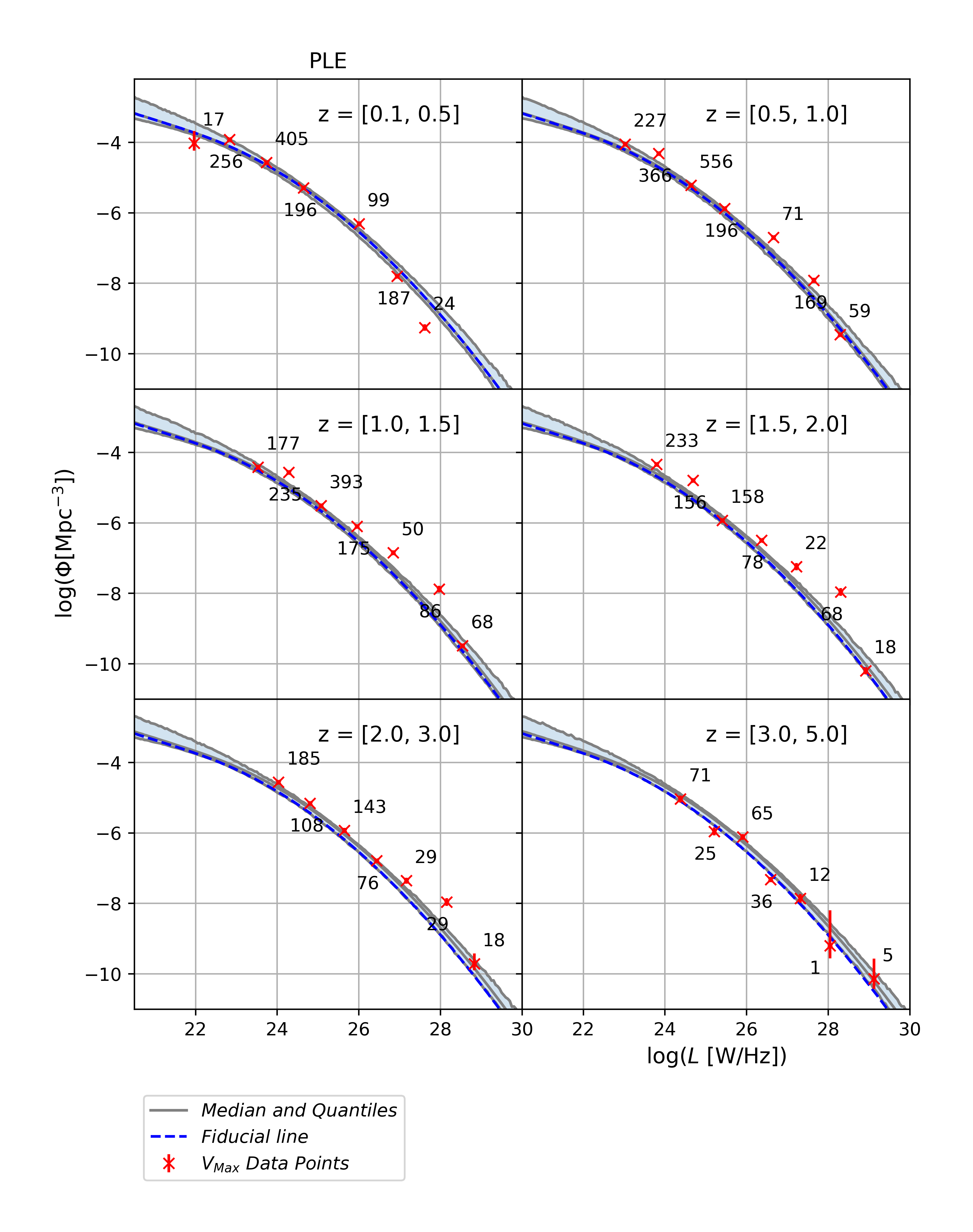}
\centering
\caption{The PLE model LF fit. The notation follows Fig. \ref{fig:LFs_LDDE}.}
\label{fig:LFs_App3}
\end{figure*}

\begin{figure*}
\centering
\includegraphics[width=0.95\textwidth]{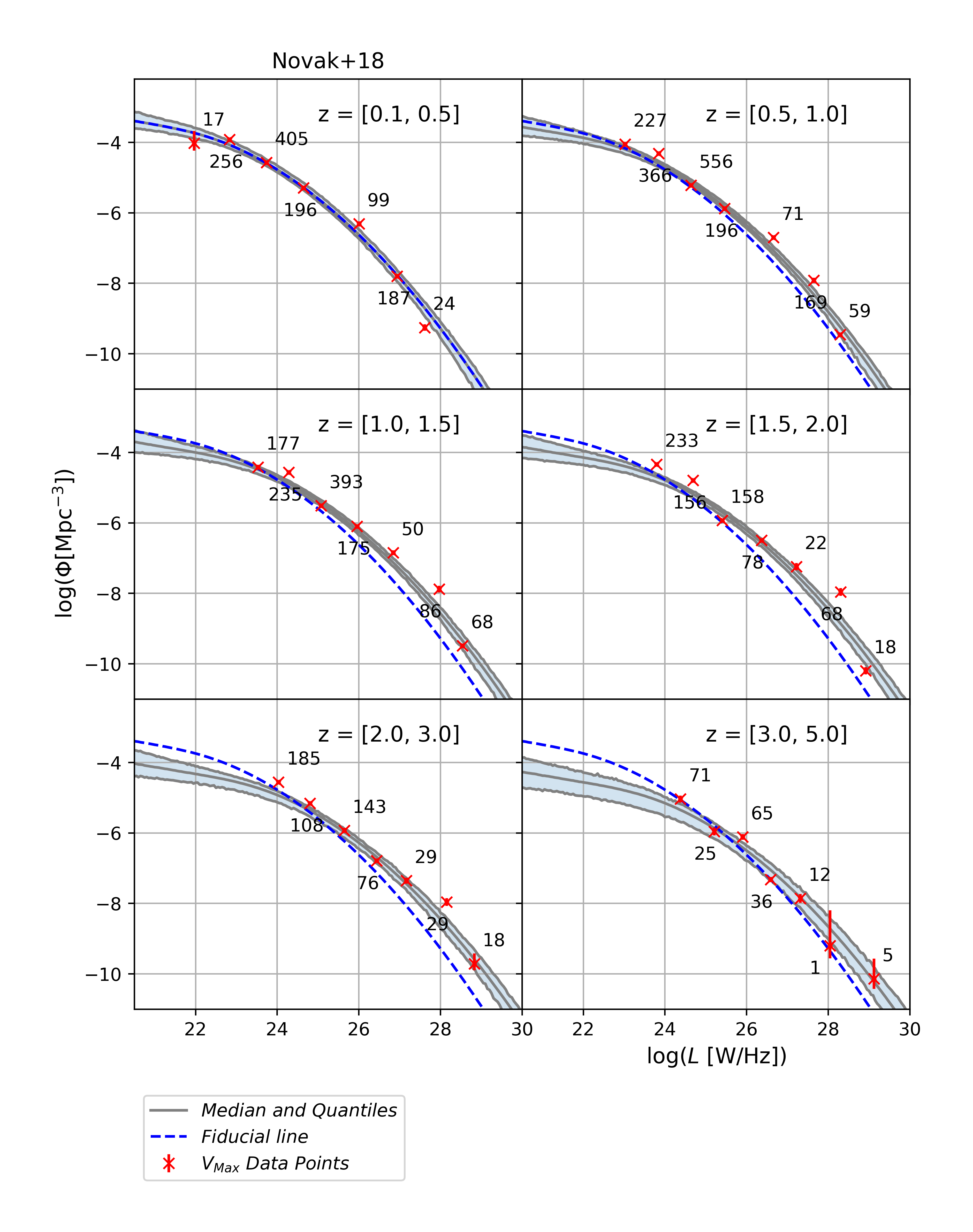}
\centering
\caption{The Novak$+18$ model LF fit. The model evolution is described in relation (\ref{eq:novakmodel}). The notation follows Fig. \ref{fig:LFs_LDDE}.}
\label{fig:LFs_App4}
\end{figure*}

\begin{figure*}
\centering
\includegraphics[width=0.95\textwidth]{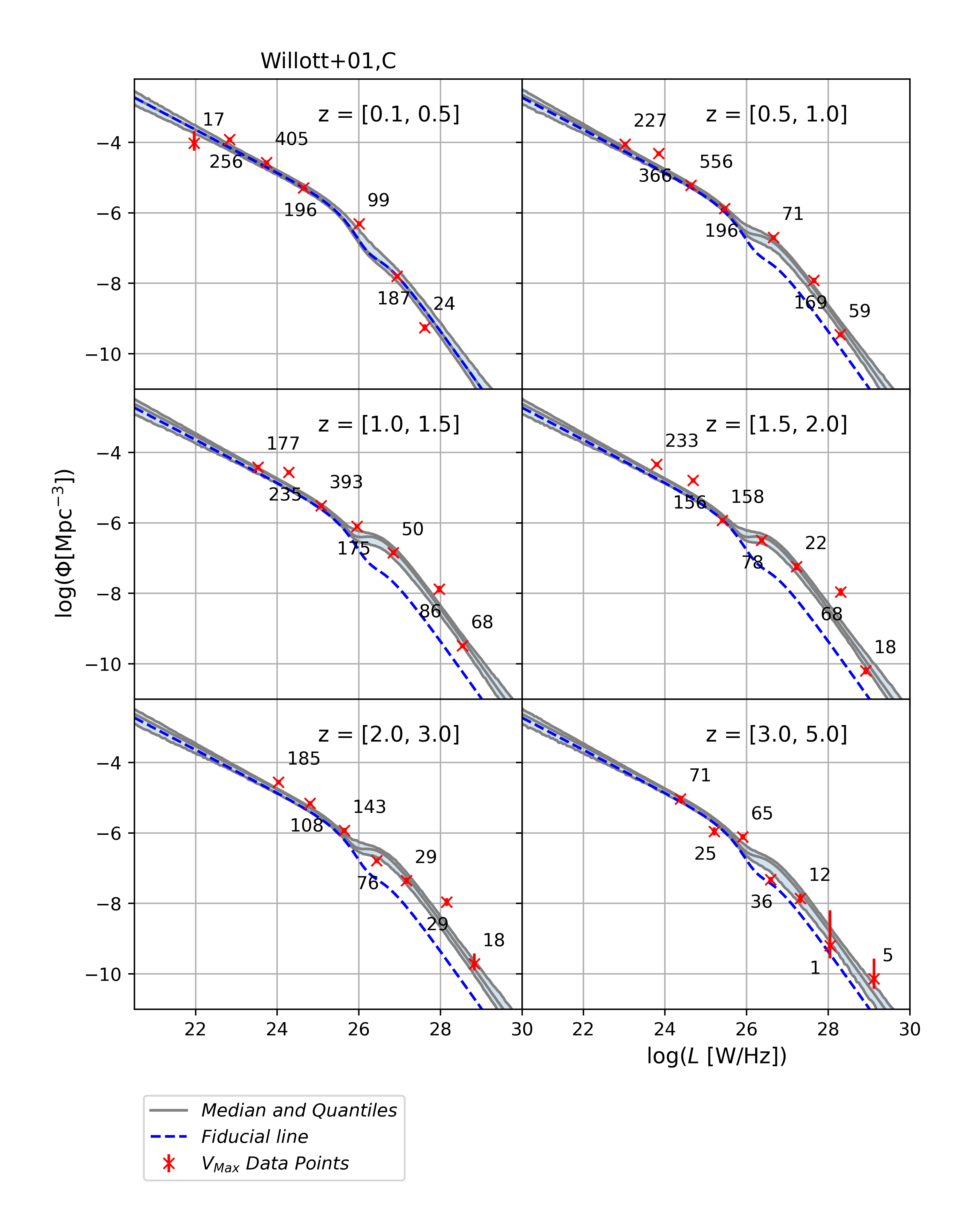}
\centering
\caption{The Willott$+01$ model LF fit. The model is described in Sect. \ref{Sect:BimodalModel}. The notation follows Fig. \ref{fig:LFs_LDDE}.}
\label{fig:LFs_App5}
\end{figure*}

\begin{figure*}
\centering
\includegraphics[width=0.95\textwidth]{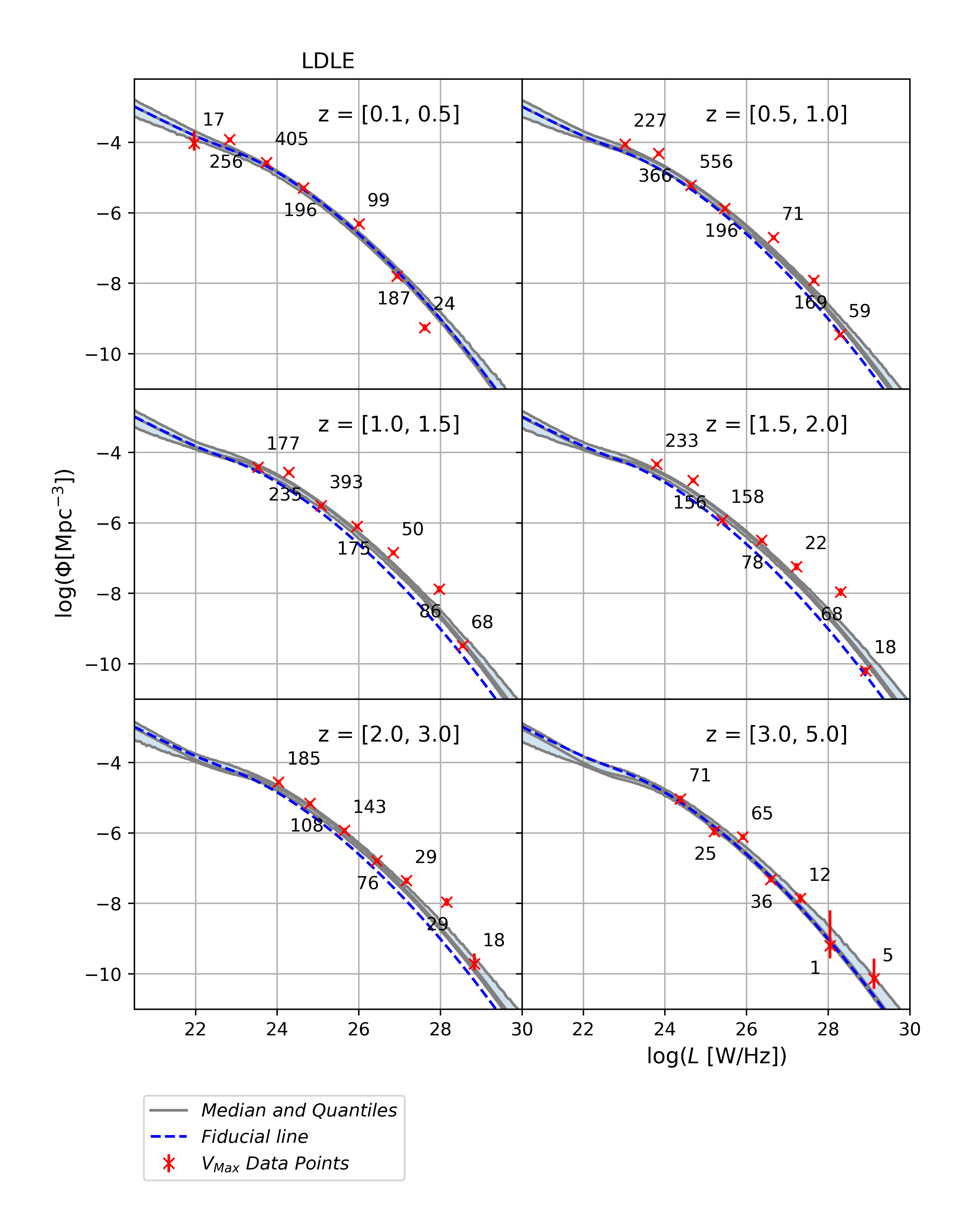}
\centering
\caption{The LDLE model LF fit. The model evolution is described in relations (\ref{eq:LDLE_Lstar}) and (\ref{eq:LDLE_ztop}). The notation follows Fig. \ref{fig:LFs_LDDE}.}
\label{fig:LFs_App6}
\end{figure*}

As described in Sect. \ref{sec:REs}, the best fitting model according to all selection criteria was the LDDE model (see Tab. \ref{Tab:ModelCompare}), described by relations (\ref{eq:LDDE}) and (\ref{eq:La}). The LDDE model LFs fit is shown in Fig. \ref{fig:LFs_LDDE}. For completeness, we show here the LF fits for all the other models described in Sect. \ref{sect:Models}, which were deemed a less optimal fit than the LDDE model. In Figs. \ref{fig:LFs_App1} to \ref{fig:LFs_App6}, we show the models along the data points obtained by using the maximum volume method. All the fits were performed on the same composite survey data set, described in Sect. \ref{sec:Data}. The LF plots were created as in Fig. \ref{fig:LFs_LDDE}.

\bibliographystyle{aa}
\bibliography{xxlpapers-bis,Refs}

\end{document}